\documentclass[10pt,journal]{IEEEtran}

\usepackage{amsmath,amssymb}
\usepackage{microtype}
\usepackage{graphicx}
\usepackage{rotating}
\usepackage{tikz}
\usepackage{tikz-cd}
\usetikzlibrary{arrows.meta, positioning, shapes.geometric, fit, calc, backgrounds,shadings}
\usepackage{helvet}
\usepackage{tabularx}
\usepackage{array}
\usepackage{booktabs}
\usepackage{multirow}
\usepackage{caption}
\usepackage{subcaption}
\usepackage{float}
\usepackage{placeins}
\usepackage{etoolbox}
\usepackage{balance}
\usepackage[hidelinks]{hyperref}

\newtheorem{researchquestion}{RQ}

\makeatletter
\def\bstctlcite#1{\@bsphack\@for\@citeb:=#1\do{\immediate\write\@auxout{\string\citation{\@citeb}}}\@esphack}
\makeatother

\newcommand{\suppref}[1]{\ref{#1}}
\newcommand{\mainref}[1]{\ref{#1}}

\begin{document}
\bstctlcite{supplementBSTcontrol}

\newcommand{\tf}[1]{_{\scriptscriptstyle\mathrm{#1}}}
\newcommand{\sysname}{CUSTOS}
\newcommand{\rqref}[1]{RQ~\ref{#1}}
\title{\sysname: Toward Forensic-Ready Zero Trust at the Capture--Containment Boundary}

\author{%
\begin{tabular}{c@{\hspace{2.5em}}c}
Avinash~Srinivasan & John~Paramadilok \\
{\small Department of Cyber Science} & {\small Joint Cyber Center} \\
{\small United States Naval Academy} & {\small U.S. Space Command} \\
{\small Annapolis, MD, USA} & {\small Peterson SFB, CO, USA} \\
{\small\texttt{srinivas@usna.edu}} & {\small\texttt{john.paramadilok@usspacecom.mil}}
\end{tabular}%
\thanks{This work has been submitted to the IEEE for possible publication. Copyright may be transferred without notice, after which this version may no longer be accessible.}%
}

\IEEEaftertitletext{%
\begin{center}
\footnotesize\emph{Disclaimer:} The views expressed in this article are those of the author(s) and do not reflect the official policy or position of the U.S. Naval Academy, Department of the Navy, the Department of War, or the U.S. Government.
\end{center}
}

\maketitle

\begin{abstract}
Zero Trust (ZT) replaces implicit trust with continuous verification, but automated containment can destroy volatile evidence before preservation. We propose \sysname, a forensic-ready ZT reference architecture whose \emph{Forensic Management Point} (FMP) links identity and policy context to tiered, rate-limited capture and orders volatile-state acquisition ahead of defender-routed destructive containment. In a controlled real-container experiment, the planted artifact was lost in all 1000 trials when capture and SIGKILL began concurrently, showing that the direct kill outran the evaluated acquisition path. The sequencing barrier completed capture before releasing that same kill in all 1000 trials. In a matched four-condition comparison, only sequencing recovered the transient artifact (200/200); a periodic snapshot-chain baseline recovered long-lived evidence (200/200) but missed the transient artifact at both cadences. Sequencing added 0.140\,s of containment delay and 9.99\,MB per event. At a 2\,s cadence, the chain added no containment delay but suspended the workload for 4.9\% of wall-clock and accrued 59.2\,KB/s after its root snapshot. The always-on decision record reduced in-process request-path throughput by 1.9--3.0\%. In-kernel enforcement and adversarial self-destruction bypass the sequencing barrier; CUSTOS preserves volatile state otherwise lost to defender-routed containment at measured cost.

\end{abstract}

\begin{IEEEkeywords}
Zero Trust Architecture, zero trust reference architecture, digital forensics, forensic readiness, zero trust forensics, incident response, identity-centric security, evidence provenance, chain of custody, volatile memory forensics, capture--containment race.
\end{IEEEkeywords}

\section{Introduction}
\label{sec:intro}
Zero Trust (ZT) has progressed from an emerging philosophy~\cite{kindervag2010,ward2014beyondcorp} to an established enterprise security model. Executive Order~14028 and OMB M-22-09 set federal ZT implementation requirements across identity, devices, networks, applications, and data~\cite{eo14028,omb2022m2209}. NIST SP~800-207 defines the Policy Decision Point (PDP) and Policy Enforcement Point (PEP), and SP~800-207A extends the model to cloud-native applications~\cite{nist800207,nist800207a}. This model replaces implicit location-based trust with continuous, identity- and policy-driven verification.

This shift changes what investigators can observe and preserve. Conventional workflows often assume that compromised systems persist, network payloads remain inspectable, and hostnames or IP addresses remain stable enough for attribution. Cloud-native ZT deployments challenge these assumptions: automated remediation may terminate ephemeral workloads before acquisition, mutual TLS (mTLS) obscures payloads from passive collectors, and identity-centric authorization distributes attribution across subjects, devices, claims, policy decisions, and short-lived sessions.

Valid-identity abuse can trigger containment before acquisition, removing memory, ephemeral state, and session or flow context. Because the activity appears authorized, recognition may require correlating records from several systems. In SolarWinds-related activity, forged SAML tokens allowed authentication as arbitrary users; CISA guidance recommended correlating cloud sign-ins with federation and domain-controller events to detect them~\cite{cisa2021aa21008a,mandiant2021unc2452}. CircleCI reported theft of a valid single sign-on (SSO) session backed by multi-factor authentication (MFA); its findings drew on authentication, network, and monitoring tools, system logs, and partners' log analyses~\cite{circleci2023incident}. These cases show why preservation must precede destructive containment that could erase context needed for recognition and reconstruction.

\begin{table*}[t]
\caption{Scope of the \sysname\ architecture and evaluation. Measured values are reported in Section~\ref{sec:feasibility}.}
\label{tab:evidence-level}
\centering
\footnotesize
\renewcommand{\arraystretch}{0.94}
\begin{tabularx}{\textwidth}{|>{\raggedright\arraybackslash}p{2.7cm}|>{\raggedright\arraybackslash}X|}
\hline
\textbf{Scope category} & \textbf{What it covers} \\
\hline
\textbf{Proposed architecture} & Forensic Management Point (FMP) reference architecture: policy-triggered tiered capture, identity- and policy-linked reconstruction, telemetry orchestration, and ZT-controlled investigative access. \\
\hline
\textbf{Implemented components} & FMP capture hook on a live FastAPI PEP (Open Policy Agent, Cedar, Casbin), plus the hash-linked decision-record log, capture and sequencing test software, Checkpoint/Restore in Userspace (CRIU) capture path, detector adapters, evidence-log tooling, and an integrated k3s ordering demonstration. \\
\hline
\textbf{Measured} & Live-gateway overhead across three policy engines; CRIU full-memory process checkpoint; the real-container sequencing-barrier experiment; containment from direct SIGKILL to forced and grace-period Kubernetes eviction; adversarial self-destruction; first-attempt command-and-control (C2) socket recovery under two detectors; and post-anchor tamper detection. \\
\hline
\textbf{Replay and modeling} & Storage and full-memory rate-cap models, telemetry-schema coverage and full-memory admission ceilings across five empirical sources and one non-empirical schema reference, detector-recall sensitivity, and a decision-record-only baseline. \\
\hline
\textbf{Evaluation boundary} & Production identity provider (IdP)/multi-cloud deployment, production-load integrated container/session/microsegment capture, live microsegment packet capture, successful cross-source incident reconstruction, coordinator and evidence-path failure, and joint detector precision--recall. \\
\hline
\end{tabularx}
\vspace{-5mm}
\end{table*}

Automated ZT response therefore creates a direct conflict between containment and preservation. We call containment-induced loss of volatile state the \emph{forensic shredder effect}; it creates a forensic gap between the system's ability to stop an incident and the investigator's ability to reconstruct it. Prior systems rarely make forensic readiness a ZT control-plane function~\cite{neale2022ztdf,inukonda2023zeta,shan2026ztforensics}.

Our central claim is that forensic-ready ZT requires an always-on minimum decision record plus richer capture constrained by a measurable time window and budget. \sysname\footnote{\sysname: Latin for guardian, keeper.} addresses that requirement through the proposed \emph{Forensic Management Point} (FMP).

\noindent\textbf{Contributions.} The paper contributes the following; Table~\ref{tab:evidence-level} separately summarizes their implementation and evaluation scope.
\begin{itemize}
\setlength{\itemsep}{1pt}
\setlength{\parsep}{0pt}
\setlength{\topsep}{2pt}
  \item \textbf{A forensic-ready ZT reference architecture.} \sysname\ centers on an FMP coordinating rate-limited capture tiers, identity- and policy-linked decision provenance, telemetry orchestration, and controlled investigative access (Sections~\ref{sec:challenges}--\ref{sec:fmp-arch}).
  \item \textbf{A ZT forensic threat model.} The model links valid-identity abuse, malicious insiders, and anti-forensic actions to five threat objectives spanning evidence destruction, capture evasion, tampering, capture-budget denial of service (DoS), and repudiation (Section~\ref{subsec:threat-model}).
  \item \textbf{A measured capture--containment race.} We characterize the race across an FMP-controlled process, real-container memory capture, k3s ordering, and managed-Kubernetes containment and detection. Evidence loss follows the destruction path and speed, not nominal grace periods; a controlled real-container experiment isolates sequencing (Section~\ref{sec:feasibility}).
  \item \textbf{A matched comparison with a periodic preservation policy.} With workload, alert and containment signals, and recovery criteria fixed, we compare bare acquisition, unsequenced and sequenced FMP capture, and a periodic checkpoint baseline inspired by Forensic Snapshot Chains (FSC)~\cite{stoyanov2026fsc}. Sequenced capture recovered the transient artifact in $200/200$ trials, versus $0/200$ for bare and unsequenced capture; the chain recovered long-lived evidence but missed that artifact because it samples periodically (Section~\ref{subsec:feas-matched}).
  \item \textbf{Telemetry-schema coverage and full-memory admission bounds.} Across five replayed public datasets, identity-oriented telemetry populated $64$--$75\%$ of the decision-record schema, versus $18$--$30\%$ for network-oriented telemetry. The coverage metric characterizes schema support, while a rate cap bounds full-memory admission (Section~\ref{sec:feasibility}).
\end{itemize}

\section{Background and Motivation}
\label{sec:bg}

\subsection{Research Questions and Evaluation Framework}
This paper addresses three research questions:
\begin{researchquestion}\label{rq:main}
How much of the FMP decision-record schema can identity- and network-oriented telemetry populate, as a bound on the context available for forensic \emph{fidelity} and \emph{correlation}?
\end{researchquestion}
\begin{researchquestion}\label{rq:adversarial}
When can destructive containment or adversarial self-destruction preempt volatile-evidence capture, and how does capture-before-containment sequencing compare with periodic checkpointing?
\end{researchquestion}
\begin{researchquestion}\label{rq:feasibility}
At the evaluated operating points, what request-path overhead, storage footprint, and full-memory admission ceiling result from tiered capture and the full-memory rate cap?
\end{researchquestion}

\rqref{rq:main} distinguishes \emph{fidelity}, whether a record preserves sufficient state and metadata, from \emph{correlation}, whether records retain shared identity, device, session, trace, and decision identifiers plus temporal-order resolution. It uses \emph{telemetry-schema coverage} only as a proxy for available reconstruction context: coverage measures source-field presence across identity, identity attributes, action, resource, risk context, verdict, and temporal-order resolution. Correlation credit uses identity and temporal-order resolution; device, session, trace, and decision identifiers are unscored. It characterizes schema support rather than joinability or incident reconstruction. \emph{Availability} (capture, retention, and accessibility) is evaluated separately because it is not a schema property.

\rqref{rq:adversarial} uses capture time $t_C$, evidence-destruction time $t_R$, and evidence-loss rate $L=\Pr[t_C\geq t_R]$. The two times are measured from the instant the response is triggered. $t_C$ is the elapsed time to capture completion; on the evaluated CRIU path, completion means that a closed local acquisition artifact exists, while durable evidence-store commit lies outside that interval. In turn, $t_R$ is the elapsed time until containment or adversarial action makes the target evidence unavailable to the capture path.

\rqref{rq:feasibility} quantifies the cost and capacity limits of tiered capture at the evaluated operating points.

\noindent\textbf{Scope and assumptions.}\label{subsec:scope}
We assume a hybrid enterprise in which perimeter evidence is supplemented by host and endpoint artifacts. We focus on four forensic assumptions challenged by cloud-native ZT deployments: centralized visibility, stable identifiers, persistent workload state, and inspectable packet content. The vendor-neutral model follows NIST SP~800-207 and SP~800-207A~\cite{nist800207,nist800207a}, with an identity provider (IdP), PDP, PEPs, and telemetry system. The evaluation assesses the architecture's capture--containment, decision-record, capacity, and telemetry properties through microbenchmarks, a live FastAPI enforcement gateway, analytical models, and schema projections over replayed telemetry datasets. Production IdP integration and end-to-end multi-cloud telemetry are planned deployment-scale extensions (Section~\ref{subsec:feas-threats}). Appendix~\suppref{sec:s-legal} maps authenticated provenance, detectable alteration, documented custody, reproducible processing, and preservation of relevant evidence to applicable evidentiary rules and ISO/IEC standards. Admissibility is case-specific; the mapping provides technical support, while production cryptographic validation is deployment-specific.

\vspace{-2mm}
\subsection{Threat Model and Security Goals}
\label{subsec:threat-model}
\noindent\textbf{System, assets, and trust.}
Protected requests traverse a PEP and are adjudicated by a PDP using identity, device, resource, and risk context. High-risk signals may trigger session termination, credential revocation, quarantine, or isolation. The FMP adds a routine decision record, targeted session/flow artifacts, and rate-limited full-state capture. The security goals are to preserve volatile incident state, retain decision provenance, and detect alteration of anchored evidence. The trusted computing base comprises the FMP capture path and evidence-store root, including its signing service, key, and periodic anchor. Workloads, requesters, the network, and privileged insiders, including operational administrators and investigators, are untrusted. This boundary is custodial rather than organizational: privilege over the monitored environment does not imply control of the evidence root. The evaluated prototype co-locates the capture hook with the PEP (Section~\ref{subsec:feas-threats}). A compromised PEP, PDP, or telemetry source can suppress or forge its signals but, by assumption, cannot control the evidence root or modify anchored records undetected. Each signed anchor covers the preceding record prefix; later records become covered by the next anchor. Under uncompromised signing and evidence-root services, anchoring detects alteration of recorded content but does not establish its semantic truth. Policy and detector correctness are separate from this integrity guarantee.

\noindent\textbf{Adversaries and threat objectives.}
We distinguish two actor classes and one anti-forensic capability. A1 is a foothold attacker with a stolen credential or compromised host; A2 is a malicious insider; and A3 represents anti-forensic actions available to either actor, including inducing rapid containment or destroying local state. Across these cases, the threat objectives are response-induced evidence destruction (TO1), evasion of richer capture (TO2), record forgery or tampering (TO3), capture-budget denial of service (DoS) (TO4), and repudiation (TO5). The timing threat occurs when $t_C\geq t_R$, and A3 actions can reduce $t_R$; successful capture is therefore a precondition for the integrity guarantees above. The gateway writes the routine record inline outside the monitored workload, protecting it from workload self-destruction. Queued persistence creates a remaining durability interval: a gateway crash before export can prevent durable storage. Targeted capture is asynchronous, while full-state capture has the greatest exposure to the timing race. Section~\ref{sec:feasibility} evaluates CRIU memory/task checkpoints, writable-area exports, response profiles, and self-destruction; future work will integrate live microsegment capture. Table~\suppref{tab:s-threat-summary} maps these cases and objectives (TO\#) to the security goals (G\#) stated next.

\noindent\textbf{Security goals and boundaries.}
The threat model yields five security goals for evidence preservation and bounded operation. The FMP targets capture before defender-routed destructive containment (G1), preservation of available decision provenance (G2), integrity and authenticated provenance (G3), binding of evidence to available identity context (G4), and bounded operational cost (G5). The evaluation demonstrates G1 on the integrated FMP-controlled k3s path and evaluates the same ordering property in controlled container-memory experiments. Production-scale assurance is planned future work. Table~\ref{tab:evidence-level} summarizes the implementation and evaluation support for G2--G5. The security claims treat policy and detector outputs as inputs and assume non-colluding authorization roles and an uncompromised evidence root. Side-channel, physical, and supply-chain attacks require separate analysis. Table~\suppref{tab:s-threat-map} maps the threat objectives to FMP mechanisms and supporting evidence.

\vspace{-2mm}
\subsection{Forensic Evidence and Readiness in ZT}
\label{subsec:identity-ecosystem}
\noindent\textbf{Identity context and forensic lifecycle.}
A ZT decision draws on principals, authentication assertions, device posture, claims, workload identity, and delegation. Attribution and reconstruction also depend on the policy version, request trace, enforcement point, and data classification. Following NIST SP~800-86's collection, examination, analysis, and reporting lifecycle~\cite{nist80086}, a ZT forensic workflow supplements disk and packet evidence with workload snapshots, IdP records, policy decisions, enforcement events, and session or flow metadata.

\noindent\textbf{Forensic readiness.}\label{subsec:forensic-readiness}
Forensic readiness requires preserving volatile evidence and correlation identifiers before defender-routed destructive containment. Under continuous verification, reconstruction depends on persistent, correlatable identifiers and the attributes used in each identity-centric policy decision. Microsegmentation makes records from multiple enforcement points necessary, while least privilege calls for just-in-time investigative access. A PDP risk or forensic-interest signal may trigger preservation before incident confirmation based on posture, behavior, privilege, or sensitive-export indicators, consistent with ISO/IEC~27043's anticipatory-collection objective~\cite{iso27043}. These requirements motivate \rqref{rq:main} and \rqref{rq:adversarial}: encryption or incomplete telemetry can reduce fidelity, fragmented identifiers impede correlation, and termination creates the capture race. Section~\ref{sec:fmp-arch} introduces \sysname, a forensic-ready ZT reference architecture centered on the FMP, to address these preservation and correlation requirements.

\section{Related Work}
\label{sec:related}

\subsection{ZT Guidance and Security Telemetry}
ZT research and federal guidance establish foundations for architecture, access control, deployment, visibility, and coordinated response~\cite{buck2021zerotrustreview,ma2025ims,nsa2024pillar}. \sysname\ extends this foundation by treating volatile-evidence preservation, identity-linked provenance, and custody as ZT control-plane objectives. Security information and event management (SIEM) and security orchestration, automation, and response (SOAR) systems coordinate detection and response~\cite{hassan2025zenguard}. ERINYES partitions serverless function activity by incoming request and combines audit and network logs into request-level provenance graphs~\cite{xi2026erinyes}. It supports post-event reconstruction, while \sysname\ coordinates volatile-state preservation before destructive ZT response. 

Host-provenance systems reconstruct attacks from audit-derived graphs~\cite{hossain2017sleuth,alsaheel2021atlas}; HADES traces identity-linked activity across machines~\cite{liu2026hades}, while tamper-evident trees protect history~\cite{crosby2009tamperevident}. At the ZT control plane, \sysname\ complements host and workload lineage with subject, device, policy, and enforcement context.

Table~\ref{tab:related-compare} compares evaluation and coordination across the forensic-readiness and acquisition approaches discussed below.
\begin{table*}[t]
\caption{Evaluation and coordination properties reported for the compared approaches.\\ \checkmark: directly reported; $\sim$: partial, indirect, or simulated; \textemdash: none identified. DFIR is digital forensics and incident response.}
\label{tab:related-compare}
\centering
\scriptsize
\setlength{\tabcolsep}{4pt}
\begin{tabularx}{\textwidth}{|>{\raggedright\arraybackslash}p{4.0cm}|*{6}{>{\hsize=.85\hsize\centering\arraybackslash}X|}>{\hsize=1.9\hsize\centering\arraybackslash}X|}
\hline
\textbf{Approach} & \textbf{Quant.\ eval.} & \textbf{Acq.\ latency} & \textbf{Freeze / disruption} & \textbf{Storage / retention} & \textbf{Request-path overhead} & \textbf{Containment coordination} & \textbf{Capture--containment boundary} \\
\hline
Neale et al.~\cite{neale2022ztdf} & \textemdash & \textemdash & \textemdash & \textemdash & \textemdash & \textemdash & \textemdash \\
\hline
ZETA~\cite{inukonda2023zeta} & $\sim$ & \textemdash & \textemdash & \textemdash & \textemdash & \textemdash & \textemdash \\
\hline
ZTForensics~\cite{shan2026ztforensics} & $\sim$ & \textemdash & \textemdash & \textemdash & $\sim$ & \textemdash & \textemdash \\
\hline
ConPoint~\cite{gharaibeh2024conpoint} & \checkmark & \checkmark & \checkmark & \textemdash & \textemdash & \textemdash & \textemdash \\
\hline
FSC~\cite{stoyanov2026fsc} & \checkmark & \checkmark & \checkmark & \checkmark & \textemdash & \textemdash & \textemdash \\
\hline
SecLaaS~\cite{zawoad2016towards} & \checkmark & \textemdash & \textemdash & \checkmark & $\sim$ & \textemdash & \textemdash \\
\hline
eBPF-for-DFIR~\cite{jin2025ebpf} & \checkmark & \textemdash & \textemdash & \textemdash & $\sim$ & \checkmark & $\sim$ \\
\hline
\textbf{\sysname\ (ours)} & \checkmark & \checkmark & \checkmark & \checkmark & \checkmark & \checkmark & \checkmark \\
\hline
\end{tabularx}
\vspace{-5mm}
\end{table*}

\subsection{Cloud Forensic Acquisition and Preservation}
Zero Trust Digital Forensics treats tools, sources, and intermediate conclusions as untrusted, motivating externalized records, provenance, and separation of duties~\cite{neale2022ztdf}. Among forensic-ready ZT proposals, Alnajjar et al.\ present a two-tier architecture for evidence collection under industrial IoT resource constraints~\cite{alnajjar2025}; ZETA assesses network-focused forensic readiness~\cite{inukonda2023zeta}; and ZTForensics demonstrates a functional prototype coupling centralized ZT enforcement with tamper-evident decision logging~\cite{shan2026ztforensics}. Container checkpoint research addresses acquisition. ConPoint collects CRIU container checkpoints for memory forensics and measures acquisition latency~\cite{gharaibeh2024conpoint}. Forensic Snapshot Chains (FSC) captures high-frequency container snapshots at a \emph{fixed cadence} after an alert and reconstructs incident timelines from the retained chain, with measured performance and storage overhead~\cite{stoyanov2026fsc}. Jin et al.\ couple continuous eBPF telemetry with automated Pod quarantine, replacement, and memory and volume acquisition~\cite{jin2025ebpf}. Their design coordinates acquisition with non-destructive quarantine; \sysname\ evaluates the timing boundary when defender-controlled response destroys volatile state. Anti-forensic research further shows that malicious software can disrupt memory acquisition by exploiting architectural features~\cite{zhang2018memoryforensic}. \sysname\ couples decision logging with tiered volatile-state preservation and characterizes capture--containment timing for defender-controlled destructive response.

Cloud-forensics research addresses architecture, collection, integrity, and access constraints. NIST's cloud forensic reference architecture defines forensic readiness for cloud deployments~\cite{nist800201}; \sysname\ couples preservation to ZT decisions and measures capture against containment. Adaptive-collection research examines selective rather than exhaustive gathering~\cite{pasquale2016adaptive}, formally derived minimum preservation requirements~\cite{alrajeh2017evidence}, and adaptive observability for evidence that disappears when containers terminate~\cite{monteiro2023adaptive}. The capture--containment boundary raises a distinct timing question: whether volatile acquisition completes before a destructive deadline. SecLaaS protects durable logs~\cite{zawoad2016towards}, while cloud-acquisition studies analyze trust dependencies across provider layers~\cite{dykstra2012cloud}. Managed-container evaluations show how lower-access deployment models limit access to critical forensic artifacts~\cite{schmid2025limits}. LiveForen attests acquisition platforms and watermarks the forensic data stream to localize in-transit tampering~\cite{liu2019liveforen}. LiveForen secures the credibility of collected evidence; \sysname\ coordinates preservation across IdP, PDP, PEP, and workload layers so volatile state remains available for collection.

\vspace{-2mm}
\subsection{Research Gap Addressed}
Table~\ref{tab:related-compare} shows a ZT design gap: other systems reporting
acquisition latency do not coordinate containment; the one coordinating it reports the
capture--containment boundary only partially. \sysname\ defines a forensic-ready ZT
reference architecture centered on an FMP that binds preservation to identity and policy
context, admits tiered capture under explicit budgets, and, when required by policy, gates
defender-routed containment on capture completion or window expiry. The evaluation compares
reactive, periodic, and sequenced policies on recovery and operational costs.

\section{Forensic Challenges in Zero Trust}
\label{sec:challenges}
Cloud-native architectures under ZT challenge three conventional forensic assumptions: persistent workloads, inspectable network payloads, and stable network identifiers.

\subsection{Ephemeral Workloads and Volatile Evidence}
Cloud-native applications run in short-lived containers and serverless instances. Volatile state can disappear through lifecycle transitions or automated response, while records remain distributed across control-plane and workload components.

\begin{itemize}
\setlength{\itemsep}{1pt}
\setlength{\parsep}{0pt}
\setlength{\parskip}{0pt}
\setlength{\topsep}{2pt}
\setlength{\partopsep}{0pt}
  \item \textit{Automated response and evidence loss.} A PDP decision or detector alert may initiate a workflow that isolates and replaces a suspicious workload. The response can stop execution and make locally held memory, command history, caches, active connections, and in-memory keys unavailable. In the measured profiles, orchestrator actions allowed the evaluated capture path to complete; direct or in-kernel termination and workload self-destruction did not (Section~\ref{sec:feasibility}).
  \item \textit{Fragmented evidence.} In serverless runtimes, memory may contain decrypted payloads, credentials, or session keys for only one invocation; container-local memory disappears when an instance terminates~\cite{gharaibeh2024conpoint,stoyanov2026fsc,monteiro2023adaptive}. Network interfaces and flow state may disappear during scaling, and tokens may expire or change during reauthorization. A single action may span IdP, PDP, PEP, and application components. Its evidence is distributed across identity records, policy decisions, enforcement logs, and transient workloads. Reconstruction must tolerate missing context and clock uncertainty.
\end{itemize}

\subsection{Encrypted Traffic and Acquisition Points}
ZT service-to-service deployments commonly use mTLS, and applications may add application-layer encryption (ALE). These controls protect confidentiality but hide application headers and bodies from passive collectors; timing, endpoints, and volume remain visible. Transport endpoints may hold plaintext in memory, whereas ALE can move plaintext exposure to a downstream application component.

\begin{itemize}
\setlength{\itemsep}{1pt}
\setlength{\parsep}{0pt}
\setlength{\parskip}{0pt}
\setlength{\topsep}{2pt}
\setlength{\partopsep}{0pt}
  \item \textit{Forward secrecy and key location.} TLS~1.3 (EC)DHE and PSK+(EC)DHE handshakes provide forward secrecy; PSK-only mode does not~\cite{rfc9846}. Obtaining a long-term authentication key later does not, by itself, enable decryption of a forward-secret session. Retrospective TLS decryption requires recorded ciphertext and traffic secrets exported during the connection or recovered from volatile endpoint state. Those secrets enable recovery of transport plaintext, but ALE-protected payloads remain opaque until application-layer decryption. Key storage in a key-management service or hardware security module does not by itself identify where plaintext exists; acquisition must target a component that processes the plaintext.
  \item \textit{Inspection trade-offs.} TLS-inspection middleboxes can restore transport-layer visibility for traffic they terminate, but not for payloads under independent end-to-end ALE. They also concentrate decryption capability and key material in a privileged component and add decryption and re-encryption overhead. The appropriate acquisition point therefore depends on whether transport termination exposes the required content or ALE defers plaintext to the application.
\end{itemize}

\subsection{Identity-Centric Attribution}
Transient addressing makes IP addresses weak attribution anchors, so reconstruction must combine identity, device, certificate, token, and policy context. This identity-centric model improves policy precision while raising four reconstruction challenges:
\begin{enumerate}[\setlength{\IEEElabelindent}{0pt}\setlength{\labelsep}{0.3em}]
\setlength{\itemsep}{0pt}
\setlength{\parsep}{0pt}
\setlength{\parskip}{0pt}
\setlength{\topsep}{2pt}
  \item \emph{Identity translation across boundaries.} OAuth token exchange supports impersonation and delegation~\cite{rfc8693}. SAML mappings, service-mesh identities, cloud roles, and successive service accounts may obscure the initiating identity in downstream records; reconstruction must preserve the subject and any available actor chain.
  \item \emph{Identity-to-location binding.} Activity may span IPs, services, regions, and policy boundaries. Binding \texttt{10.1.5.20} to \texttt{User~A} at a specific instant requires time-resolved identity, session, gateway, and address-allocation records with synchronized clocks and sufficient timestamp precision.
  \item \emph{Successive authorization decisions.} Reauthentication and policy re-evaluation need not open new sessions and may produce successive decisions that investigators must correlate using token, trace, and decision identifiers.
  \item \emph{Service identities and actor attribution.} A stolen service credential can produce well-formed API calls that resemble legitimate automation. Application and cloud audit records may retain the actions, but reconstruction must distinguish malicious use from authorized activity under the same principal.
\end{enumerate}

Together, these challenges require preservation before termination, acquisition where plaintext is processed, and correlation across identity, policy, enforcement, and workload records.

\begin{figure*}[t]
\centering
\sffamily
\resizebox{\textwidth}{!}{\input{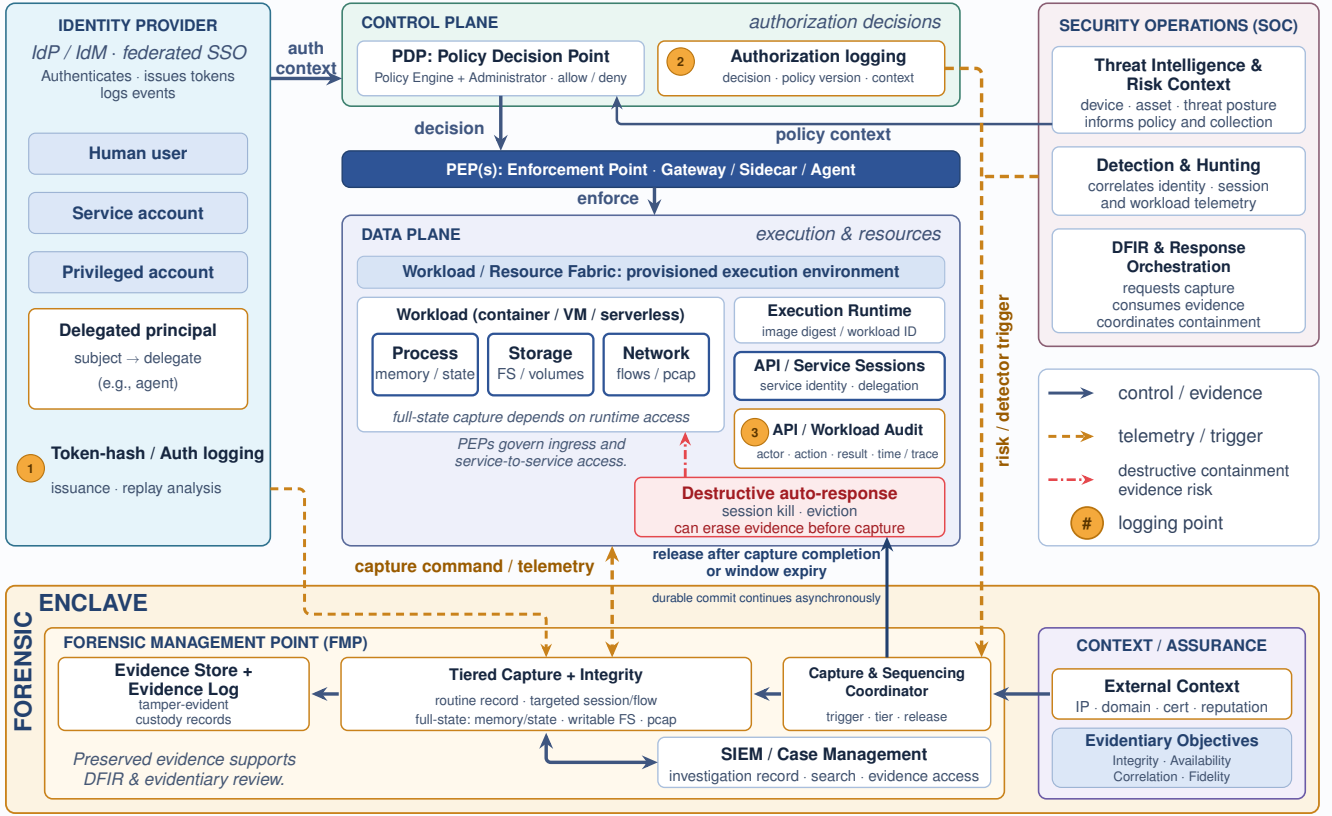}}
\caption{\sysname\ logical forensic-ready ZT reference architecture across identity, control, enforcement, and data planes. Under the default release policy, the FMP delays defender-routed destructive containment until capture completion or capture-window expiry; durable evidence-store commit continues asynchronously.}
\label{fig:fmp-zta-column}
\vspace{-5mm}
\end{figure*}

\section{\sysname\ Reference Architecture}
\label{sec:fmp-arch}
\sysname\ addresses these requirements through the logical reference architecture shown in Figure~\ref{fig:fmp-zta-column}; Section~\ref{sec:feasibility} evaluates its core preservation mechanisms and capture--containment properties in the prototype. The architecture uses identity- and policy-rich ZT telemetry for four forensic functions. \emph{Policy-triggered state capture} preserves volatile evidence before destructive containment. \emph{Identity- and policy-linked reconstruction} associates events with authenticated identity and available device and policy context rather than transient IP addresses. \emph{ZT-controlled investigative access} applies the same policy discipline to evidence use. \emph{Telemetry orchestration} coordinates identity, workload, and enforcement records in an integrity-protected evidence stream.

\subsection{FMP Placement and Interfaces}
The FMP receives identity, policy, enforcement, and detection inputs through control-plane interfaces and coordinates capture before defender-routed destructive remediation. The PDP's access decision proceeds independently of capture; only the subsequent remediation path is sequenced. Routine records remain inline, with their request-path latency measured in Section~\ref{sec:feasibility}. \emph{Inputs} include PDP verdicts and policy versions, PEP enforcement events, IdP claims, detector or forensic-interest signals, and workload/session metadata. \emph{Outputs} include tiered PEP/runtime capture commands, integrity-protected evidence bundles, per-decision provenance records, and just-in-time investigator-access decisions. These interfaces support capture sequencing and cross-source case assembly.

\subsection{Policy-Triggered State Capture}
When a risk or forensic-interest signal meets a preservation-policy trigger, the FMP initiates state capture. It selects a tier and invokes PEP, runtime, or SOAR actions to preserve workload memory, authorized session or microsegment traffic, and ephemeral logs before destructive response makes them unavailable.

\subsubsection{Tiered Capture at Scale}
To scale preservation independently of routine decision volume, the FMP applies three capture tiers:
\begin{enumerate}[\setlength{\IEEElabelindent}{1.25em}\setlength{\labelsep}{0.3em}]
\renewcommand{\labelenumi}{\roman{enumi}.}
\setlength{\itemsep}{0pt}
\setlength{\parsep}{0pt}
\setlength{\parskip}{0pt}
\setlength{\topsep}{2pt}
  \item \emph{Routine decision record} (always-on): records the action, resource, verdict, policy version, request trace ID, and available authenticated identity and associated device context for each decision.
  \item \emph{Targeted preservation} (event-triggered): asynchronously preserves validated identity-claim metadata, a token hash rather than the bearer secret, request metadata, and session or microsegment flow records for requests associated with policy constraints, authentication challenges, or anomalous activity.
  \item \emph{Workload-scoped full-state capture} (selective and rate-limited): at operator-controlled runtime boundaries, preserves process memory and execution state, the writable container filesystem, and authorized packet captures for admitted high-risk events. Its defined boundary is the local workload, with sequential component acquisition rather than a mutually atomic snapshot that includes global infrastructure or remote service state.
\end{enumerate}
Here, \emph{full-state} denotes this tier, whereas \emph{full-memory} denotes its measured CRIU memory component used to size the rate cap. Every decision produces a routine record; an escalated event additionally receives either targeted or full-state capture, and an event denied full-state admission falls back to targeted preservation. Admission policies can partition the budget by identity or source to limit capture-budget DoS. Production integration of live microsegment capture and per-identity enforcement is planned; Section~\ref{subsec:feas-robustness} evaluates the admission behavior of per-identity budgeting.

\subsubsection{Detection Inputs and Capture Triggers}
The FMP is trigger-decoupled: signature-based detectors, behavioral analytics, anomaly detectors, and threat intelligence supply preservation signals. A single anomaly may trigger targeted preservation, while high-confidence or correlated signals may trigger rate-limited full-state capture. The interface accepts either an upstream risk score or a categorical signal. Deployments select the scoring method, thresholds, and capture budgets; Section~\ref{sec:feasibility} evaluates one reference operating point and quantifies sensitivity to upstream detector recall.

\subsubsection{Capture--Containment Sequencing}
\label{subsec:sequencing}
For defender-routed destructive containment, the FMP sequences capture and release to reduce the risk of $t_C \geq t_R$. Delaying destructive containment increases dwell time, while containing first may destroy volatile evidence. A workload-specific \emph{capture window} bounds this trade-off. The window opens when the destructive-containment hold becomes active, not when a signal is generated or received; it bounds held dwell time rather than detection latency. A production deployment places expiry enforcement locally at the PEP, runtime, or orchestrator so release remains independent of FMP availability (Section~\ref{subsec:feas-threats}). Sequencing follows three release rules:
\begin{enumerate}[\setlength{\IEEElabelindent}{1.25em}\setlength{\labelsep}{0.3em}]
\renewcommand{\labelenumi}{\roman{enumi}.}
\setlength{\itemsep}{0pt}
\setlength{\parsep}{0pt}
\setlength{\parskip}{0pt}
\setlength{\topsep}{2pt}
  \item \emph{Capture-complete release.} The default policy releases destructive containment when capture closes the local artifact; durable evidence-store commit continues asynchronously. This policy applies when the closed artifact survives containment, as on the evaluated workload-scoped paths. Node-loss scenarios require durable export before release. If the capture window expires first, release occurs even if capture is incomplete.
  \item \emph{Durable-commit release.} A deployment requiring durable commit before release keeps the target frozen or non-destructively isolated until the flush completes; resuming it during commit increases attacker dwell. Non-destructive controls, such as traffic mirroring or access restriction, may remain active throughout capture and commit.
  \item \emph{Containment-priority release.} High-impact resources may use a zero-length window to prioritize containment, while other workloads may accept bounded delay to preserve more evidence.
\end{enumerate}
Sequencing governs defender-routed destructive containment, while workload-local self-destruction remains within the A3 timing threat defined in Section~\ref{subsec:threat-model}. The always-on routine record provides the minimum evidence designed to survive that behavior. For observable operations such as memory acquisition or traffic mirroring, PEP, runtime, or host-local enforcement can narrow the opportunity for A3 to react; Section~\ref{sec:feasibility} measures the residual race.

\subsection{Identity- and Policy-Linked Reconstruction}
The architecture supports reconstruction by linking records from the IdP, PDP, PEP, workload runtime, and related sources rather than relying on transient IP addresses (Section~\ref{sec:challenges}). The linked records preserve source timestamps, request trace IDs, and available timing metadata so that ordering uncertainty remains explicit. They support three functions:
\begin{enumerate}[\setlength{\IEEElabelindent}{1.25em}\setlength{\labelsep}{0.3em}]
\setlength{\itemsep}{0pt}
\setlength{\parsep}{0pt}
\setlength{\parskip}{0pt}
\setlength{\topsep}{2pt}
  \item \emph{Within-session correlation.} Request trace IDs and session identifiers associate records within a session with user and device identifiers, validated claim metadata, and a token or assertion hash.
  \item \emph{Cross-session reconstruction.} Repeated authentication and authorization can distribute an intrusion path across short-lived sessions. Available identifiers in reauthentication, delegation, and policy-decision records support reconstruction of that sequence.
  \item \emph{Attribution context.} The linked context expands ``\texttt{Machine~A accessed File~B}'' into a request authenticated as Identity~X, associated with Device~Y and Certificate~Z, and evaluated under Policy~P when accessing File~B. This provides richer system-level attribution context, while human attribution requires corroborating evidence.
\end{enumerate}
Appendix~\suppref{sec:s-legal}, particularly Table~\suppref{tab:s-evidence-mapping}, maps the resulting evidence properties to applicable legal and standards criteria.

\subsection{ZT Controls for Evidence Handling}
The architecture extends ZT principles across three areas: evidence integrity and investigator access, FMP administration, and regulated-data handling.

\subsubsection{Evidence Integrity and Investigator Access}
The evidence store retains a write-protected master copy of each artifact; authorized workflows derive working copies as needed. Investigator accounts hold no standing privilege and receive expiring just-in-time access for authorized investigations. Destructive evidence actions require two-party approval. Every evidence action is recorded in the append-only evidence log. Artifact hashes are recorded at capture, and log records are hash-linked. Under the trust assumptions of Section~\ref{subsec:threat-model}, signed anchors make later alteration of covered records, including administrator actions, detectable. For scalable verification, periodically signed anchors may be aggregated in a global Merkle root (Section~\ref{subsec:feas-robustness}).

\subsubsection{FMP Administration and Security}
The FMP's policy-administration and capture privileges make it a high-value control-plane service requiring explicit protection. Capture policies, escalation thresholds, retention rules, and access grants are policy-as-code artifacts governed by the same review, version-control, and signing workflow as access policies. Its privileged operations include evidence-store writes, capture control at PEPs and runtimes, IdP session reads, and SOAR invocation. Two design requirements protect these operations. First, \emph{scope minimization} restricts each privilege: standing evidence-store access is append-only; IdP access is read-only; PEP, runtime, and SOAR invocations use narrowly scoped, short-lived service credentials; and policy modification, evidence release, and privileged administration require just-in-time, two-party grants. Second, \emph{administrative auditing} records FMP actions in the evidence log. Enclave microsegmentation and out-of-band access restrict administrative paths. Within the same trust boundary, these controls protect administrative integrity and traceability, while source semantics require independent corroboration. FMP maintenance is assigned to a distinct \emph{FMP operator} role under the same authorization workflow.

\subsubsection{Regulated Data Controls}
For evidence containing protected health information or Payment Card Industry Data Security Standard (PCI DSS) account data, the architecture associates each artifact with the applicable encryption, retention, masking, and access-control requirements using the data-classification context of Section~\ref{subsec:identity-ecosystem}. These design requirements define the planned regulated-data production integration; Appendix~\suppref{sec:s-regulated} details their policy mapping and enforcement scope.

\subsection{Telemetry Orchestration and Deployment Model}
\sysname\ extends the PDP/PEP model defined in NIST SP~800-207~\cite{nist800207} with its FMP, which coordinates forensic preservation and enforces evidence-handling policy. The FMP runs as a control-plane service within a hardened \emph{forensic enclave}. The logical model defines this enclave as an administrative and network isolation boundary; deployments may additionally use a hardware trusted execution environment. The architecture augments the PEP interface with mirror/capture functions alongside \textit{allow}/\textit{deny} enforcement. On a forensic-interest signal from the FMP, the PEP can mirror traffic authorized for forensic capture or export metadata to the enclave.

For transport-encrypted sessions, policy-authorized endpoint instrumentation may export per-session traffic secrets under hardware-backed key control, enabling decryption of the corresponding captures. Capture policy constrains the scope, retention, and access of these secrets. When ALE keys remain outside the enclave, the capture path preserves ciphertext and metadata; plaintext acquisition instead targets an authorized application component that performs decryption (Section~\ref{sec:challenges}). In both paths, evidence is staged near the acquisition point and exported to the isolated enclave for retention and controlled access.

\section{\sysname\ Feasibility and Evaluation}
\label{sec:feasibility}
We evaluate \sysname\ along three tracks: \emph{(E1)} schema projection and storage modeling; \emph{(E2)} capture, containment, robustness, and matched-baseline experiments; and \emph{(E3)} request-path overhead from an FMP hook on a live FastAPI gateway. Expensive tiers are timed separately and composed into the model; a 30-run k3s experiment demonstrates capture-before-containment ordering. Together, these tracks quantify the architecture's core capture--containment properties under controlled workloads. Appendix~\suppref{sec:s-testbed} gives the testbed and sample counts; Section~\ref{subsec:feas-threats} bounds the claims.

\subsection{Tiered Overhead Decomposition}
\label{subsec:feas-overhead}
For decision $d$ at tier $t\in\{\mathit{routine},\mathit{targeted},\mathit{full}\}$, FMP processing through queued transport comprises five latency components:
{\small
\begin{equation}
T_{\text{artifact}}(d,t) = T_{\text{trigger}} + T_{\text{capture}}(t) + T_{\text{serialize}} + T_{\text{hash}} + T_{\text{transport}}.
\label{eq:t-artifact}
\end{equation}}
$T_{\text{trigger}}$ is tier-selection latency after the decision or an upstream risk or forensic-interest signal. $T_{\text{capture}}(t)$ varies most: it is negligible for routine records and was not separately measured for the evaluated targeted-record serialization path; full-memory acquisition ranges from tens of milliseconds to seconds, depending on boundary and size. $T_{\text{serialize}}$ canonicalizes the record; the targeted path contributes its reported $\approx$0.08\,ms here. $T_{\text{hash}}$ computes and links its SHA-256 digest, and $T_{\text{transport}}$ exports it to the evidence store. Store verification and durable flush are reported separately below. Equation~(\ref{eq:t-artifact}) describes per-event processing through queued transport, not request-path delay or durable-commit time. In the prototype, the measured inline increment includes only routine-record serialization and hash linking. Targeted and full-state work executes asynchronously, and export is queued off path. Periodic anchor signing is excluded. Because the equation is an accounting decomposition, separately reported medians do not sum to it. Expensive work is thus \emph{incurred only for escalated events}: cost scales with the escalation fraction, not full-memory capture per decision.

\subsection{Cost and Storage Model}
\label{subsec:feas-cost}
The tier structure governs logical retained storage. Let $r_t$ be the artifact rate of tier $t$ (events/s), $s_t$ its average bytes per artifact, and $\rho_t$ its retention window (days). Under stationary rates and expiry, the steady-state footprint before replication, compression, and backend overhead is:
\begin{equation}
\mathit{Footprint}_{\text{ss}} \;=\; 86\,400 \cdot \sum_{t \in \mathit{tiers}} r_t \cdot s_t \cdot \rho_t.
\label{eq:storage}
\end{equation}

Every decision produces a routine record; an escalated event additionally receives either targeted or full-state capture, never both, as the spill term in Eq.~\ref{eq:spill} makes explicit. Let $R$ be the decision rate, $h$ the labeled high-risk fraction, $f_t$ the base targeted fraction excluding that high-risk stream, and $c$ the full-memory rate cap (events/s). High-risk arrivals above $c$ spill down to targeted preservation:
{\small
\begin{equation}
r_{\mathit{routine}}=R;\, r_{\mathit{full}}=\min(hR,c);\, r_{\mathit{targeted}}=f_tR+\max(hR-c,0)
\label{eq:spill}
\end{equation}}

At the reference point ($R{=}500$/s, $h{=}0.02$, $c{=}0.05$/s, $f_t{=}0.124$), $r_{\mathit{full}}{=}0.05$/s and $r_{\mathit{targeted}}{=}71.95$/s: a $62.0$/s base stream plus $\approx$9.95/s spilled from the capped high-risk stream. Equation~(\ref{eq:storage}) thus uses post-cap artifact rates, not raw high-risk arrivals. Table~\suppref{tab:s-tier-params} gives the reference per-tier rates and conservative sizes. Routine and targeted storage still scale with $R$, whereas the cap bounds the full-memory contribution to $86\,400 \cdot c \cdot s\tf{full} \cdot \rho\tf{full}$: rate limiting prevents full-memory storage from growing without bound with throughput, unlike uniform per-decision memory capture.

\vspace{-2mm}
\subsection{The Capture--Containment Race}
\label{subsec:feas-race}
Section~\ref{subsec:threat-model} defines evidence loss when $t_C \ge t_R$. Both clocks start at the shared response trigger: $t_C$ ends at capture completion, and $t_R$ ends when containment or adversarial action makes the target evidence unavailable. With tier selection fixed, the race experiments measure CRIU's $T_{\text{capture}}$ interval in Eq.~\ref{eq:t-artifact}. The same race applies to session, filesystem, and flow artifacts. Capture succeeds when
\begin{equation}
t_C \;<\; t_R.
\label{eq:race}
\end{equation}
Conditioned on tier, workload, and destruction profile, the \emph{evidence-loss rate} $L=\Pr[t_C\geq t_R]$ is the fraction of triggered events that lose the artifact and the complement of the capture-success probability in Eq.~(\ref{eq:race}). Detector latency upstream and durable commit downstream therefore lie outside $L$. It is the primary preservation metric evaluated below. The capture window of Section~\ref{sec:fmp-arch} limits defender-imposed delay; the self-destruction experiments below measure the residual adversarial race.

\vspace{-2mm}
\subsection{Measured Results}
\label{subsec:feas-results}
The evaluation directly measures canonical serialization, SHA-256 hash linking, RSA-2048 anchor operations, flow-record serialization, and memory-image hash/write throughput, alongside the live gateway (E3), runtime and orchestrator measurements, and an integrated k3s ordering demonstration. Table~\suppref{tab:s-tier-params} combines the measured operation latencies with reference rates, sizes, and retentions at 500 decisions/s, a 2\% high-risk arrival fraction capped per~(\ref{eq:spill}), and a 256\,MiB memory artifact. Absolute latencies remain host-dependent.

\begin{figure}[t]
\centering
\includegraphics[width=\columnwidth]{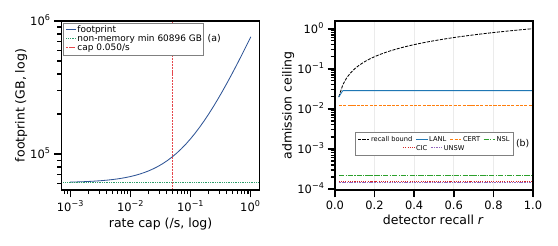}
\vspace{-8mm}
\caption{Feasibility bounds. (a) Steady-state storage vs full-memory rate cap (dashed $0.05$/s); (b) admitted malicious-event fraction $\min(r,C)$ vs.\ recall $r$ for source-specific ceiling $C$ (dashed $y{=}r$).}
\label{fig:feas-bounds}
\vspace{-4mm}
\end{figure}

\begin{figure}[t]
\centering
\includegraphics[width=\columnwidth]{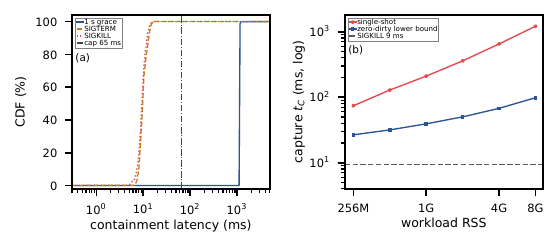}
\vspace{-8mm}
\caption{Capture--containment race. (a) Containment-latency CDFs against a 65\,ms median-capture reference; (b) $t_C$ versus resident set size, RSS (single-shot and pre-dump) against the $\approx$9\,ms kill.}
\label{fig:feas-race}
\vspace{-4mm}
\end{figure}

\begin{table}[t]
\centering
\caption{Live-PEP overhead (Zen~5, 30 paired repeats; on-minus-off latency deltas in ms). Off/on rates are marginal medians; throughput cost is the median paired change, so the two need not agree.}
\label{tab:overhead}
\scriptsize
\setlength{\tabcolsep}{1.5pt}
\resizebox{\columnwidth}{!}{%
\begin{tabular}{|l|l|r|r|r|r|r|}
\hline
\textbf{Engine} & \textbf{Deploy} & \textbf{off/on (rps)} & \textbf{Tput cost} & \textbf{$\Delta p50$} & \textbf{$\Delta p95$} & \textbf{$\Delta p99$} \\
\hline
OPA    & sidecar    & 839/839   & $1.0\%$ & $+0.08$ & $+0.39$ & $+1.93$ \\
\hline
Cedar  & in-process & 3197/3093 & $3.0\%$ & $+0.62$ & $+0.75$ & $+0.76$ \\
\hline
Casbin & in-process & 2599/2543 & $1.9\%$ & $+0.55$ & $+0.61$ & $+0.30$ \\
\hline
\end{tabular}}
\vspace{-5mm}
\end{table}

\subsubsection{Request-Path Overhead}
Microbenchmarks place canonical serialization at $\approx$6.8\,$\mu$s and a hash-chain link at $\approx$0.4\,$\mu$s; RSA-2048 signing costs $\approx$0.16\,ms but occurs only at 1000-entry anchors, off the request path. Because only the routine record is inline, its measured cost is $\approx$7.1\,$\mu$s at $p50$ and $\approx$7.4\,$\mu$s at $p95$; targeted-record serialization ($\approx$0.08\,ms) and full-memory checkpointing ($\approx$65\,ms) run asynchronously. On the live FastAPI PEP, routine recording costs $3.0\%$ (Cedar) and $1.9\%$ (Casbin) throughput for the in-process engines, with $\Delta p99\le0.76$\,ms; the OPA sidecar shows a $1.0\%$ median paired change within run-to-run noise and reaches $+1.93$\,ms at $p99$, consistent with greater sidecar and network variability (Table~\ref{tab:overhead}). The in-process $1.9$--$3.0\%$ range is therefore the more stable marginal-cost estimate.

\subsubsection{Storage Footprint}
Table~\suppref{tab:s-tier-params} gives the reference rates and conservative sizes. Measured sizes and latencies come from experiments; the conservative sizes support the $\approx$96\,TB scenario; the full-memory cap and retentions are operator-selected; and $R$, $h$, and $f_t$ are illustrative. At this operating point, conservative routine and targeted artifacts ingest $\approx$474\,GB/day. The cap bounds the full-memory contribution, so the footprint approaches the non-full-state contribution below $\approx10^{-2}$/s and becomes cap-driven above it (Fig.~\ref{fig:feas-bounds}(a)). At 365/90/30-day routine/targeted/full-memory retention, measured sizes give $\approx$46\,TB and conservative sizes $\approx$96\,TB; an 8\,GiB workload at the same cap and retention implies $\approx$1.1\,PB for full memory alone. At measured sizes, uncapped memory capture for every high-risk event or every decision would require $\approx$7.0\,PB or $\approx$352\,PB, respectively ($152\times$ and $7{,}594\times$ the tiered footprint).

\subsubsection{Destruction Speed and Release Policy}
Three boundaries organize these measurements: capture completion yields a closed local artifact; containment release permits the destructive response to proceed; durable commit verifies the artifact in the evidence store. They distinguish acquisition time, added response delay, and persistence time.
A force-kill terminated the 256\,MiB workload in $\approx$9\,ms and a cooperating SIGTERM in $\approx$10\,ms, both inside the $\approx$65\,ms capture interval ($L{=}100\%$), whereas a 1\,s configured SIGTERM grace ($\approx$1.13\,s measured) lay outside it ($L{=}0\%$; Fig.~\ref{fig:feas-race}; $n{=}1000$ each). A separate Docker/\texttt{runc} release-policy study ($n{=}1000$ per condition) evaluates three release points, all preserving the checkpoint ($L{=}0\%$). Releasing at checkpoint completion holds containment for $\approx$175\,ms and reduces a 64\,KiB \texttt{pwrite()} activity proxy by $93.0\%$ relative to an equal ungated delay because CRIU seizes the task during the dump. Releasing only after synchronous durable commit, with the workload resumed during commit, extends the hold to $\approx$1314\,ms but reduces activity by only $12.2\%$. Holding the container cgroup frozen through commit extends the hold to $\approx$1344\,ms and reduces activity by $98.1\%$. Capture completion is therefore the lowest-delay evaluated release point that preserved the checkpoint, and a policy that waits for durable commit must keep the target frozen or state-preservingly isolated through commit to prevent continued execution during commit. The sequencing result of Section~\ref{subsec:feas-matched} is size-robust from 256\,MiB to 8\,GiB ($L{=}0\%$ vs.\ $100\%$ without the hold; Table~\suppref{tab:s-sizesweep}). It reproduces on a second host (EPYC~7702) at 256\,MiB and 1\,GiB (Table~\suppref{tab:s-crosshost}), where the release ordering is invariant and only the absolute holds scale.

\subsubsection{CRIU Capture Validation}
\begin{figure}[t]
\centering
\includegraphics[width=\columnwidth]{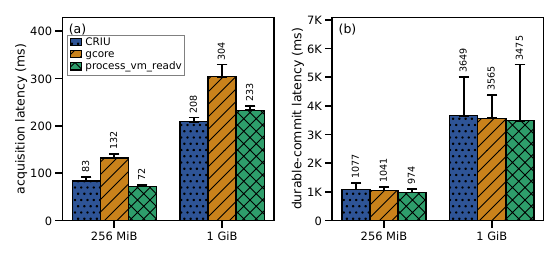}
\vspace{-8mm}
\caption{Matched-boundary capture methods (Zen~5, $n{=}200$/method/size; medians and $p95$): (a) closed local artifact; (b) verified durable commit.}
\label{fig:capture-methods}
\vspace{-5mm}
\end{figure}

We use a non-destructive CRIU checkpoint (\texttt{criu dump --leave-running}) of a 256\,MiB workload, capturing memory pages, registers, open descriptors, and task state. On the Zen~5 host, capture took $\approx$65\,ms median and $\approx$75\,ms at $p95$ ($n{=}1000$, range 54.8--114.4\,ms); an older EPYC~7702 host gave $\approx$126\,ms. Under matched timing boundaries over 128/256/512/1024\,MiB (Fig.~\ref{fig:capture-methods}(a); $n{=}200$/method/size/host), CRIU had the smallest fitted size coefficient on both hosts, while \texttt{gcore} was slowest in every configuration. Descriptive intercept-plus-size fits have $R^2\ge0.997$; coefficients, fitted crossovers, and the independent near-crossover validation are in Appendix~\suppref{sec:s-cost}. Durable-commit method spreads are smaller ($5$--$29\%$), consistent with the common persistence path dominating total latency. The matched setup equalizes timing boundaries, not invocation mechanics: CRIU and \texttt{gcore} run as subprocesses, whereas \texttt{process\_vm\_readv} executes within the test program. The matched $83$\,ms and standalone $65$\,ms CRIU results are therefore reported separately. We select CRIU for restorable memory and task-state fidelity; \texttt{gcore} yields a non-restorable core and \texttt{process\_vm\_readv} a non-atomic, memory-only acquisition without restorable task state. Race outcomes depend on workload size. Single-shot $t_C$ reaches $\approx$1.2\,s at 8\,GiB RSS, below the $\approx$2.1\,s eviction median. At zero dirty rate, pre-dump limits the 8\,GiB final freeze to $\approx$98\,ms (Fig.~\ref{fig:feas-race}(b)); this is a zero-dirty lower bound, since at 256\,MiB the freeze grows $\approx$27$\to$46\,ms as dirtying reaches 100\,MB/s.

The k3s ordering study uses an FMP-controlled process because this setup's \texttt{crictl checkpoint} interface does not expose the container address space~\cite{kep2008}. We separately validate the race at a full-container boundary by checkpointing a 260\,MiB Docker container (full address space, \texttt{--tcp-established}) with \texttt{runc}/CRIU. Its $\approx$160\,ms median ($n{=}1000$) was $\approx 2.5\times$ the Zen~5 process median but remained far above a direct container kill ($\approx$9\,ms, $n{=}200$), so $L{=}100\%$ also holds for the container.

The $\approx$65\,ms CRIU time excludes durability. A writable-area export of the container's \texttt{/root} adds $\approx$13--45\,ms, yielding an $\approx$80--110\,ms sequential, non-atomic composition; a full root-filesystem export costs $\approx$330\,ms. A local checkpoint-to-\texttt{fsync} experiment ($n{=}1000$, 256\,MiB) yielded a $\approx$1232\,ms median ($p95\approx1.66$\,s). Median checkpoint/hash/transport/store-verification/flush stages are $175/138/42/139/703$\,ms; stage medians are non-additive, and the per-trial unattributed residual has a $33$\,ms median. Falco and Suricata alert-to-test-program latencies are $\approx$0.69/$0.80$\,s, whereas Table~\ref{tab:detectors}'s $\approx$0.16/0.23\,s values are terminating-checkpoint durations. These separately measured legs characterize individual stages rather than an end-to-end detector-to-durable race.

\begin{figure}[t]
\centering
\includegraphics[width=\columnwidth]{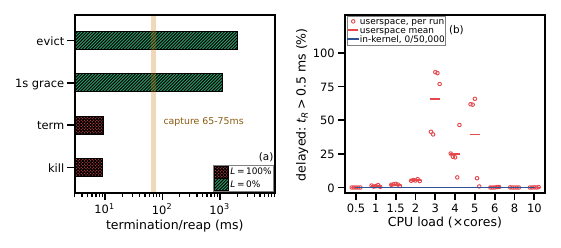}
\vspace{-8mm}
\caption{(a) Per-mechanism termination/reap $t_R$ vs.\ full-memory capture $t_C$; the grace bar is a 1\,s configured grace measured at $\approx$1.13\,s. (b) Share of enforcement observations slower than $0.5$\,ms, a threshold fixed before the data was inspected, for five independent runs per load ($n{=}1000$ per condition per run). Dashes are run means; in-kernel recorded $0/50{,}000$ delayed observations.}
\label{fig:eval-race}
\vspace{-5mm}
\end{figure}

\subsubsection{Remediation and Orchestrated Eviction}
Against the capture distribution above, container \texttt{stop}, control-plane revocation, and Kubernetes eviction leave margin ($L{=}0\%$, memory-scoped; Table~\suppref{tab:s-remediation}). On single-node k3s, forced eviction took 2060\,ms median ($p95\approx2585$\,ms) and grace expiry $\approx30$\,s ($n{=}100$ each): force removes the API object before runtime memory is reclaimed. The same timing range was observed on a three-node DigitalOcean Kubernetes Service (DOKS) cluster, where forced deletion, rolling replacement, preemption, and node \texttt{drain} fell within $\approx2.1$--$2.2$\,s and grace expiry took $\approx30.09$\,s. We do not rank mechanisms within the first band. Rolling replacement initially retained $3000/3731$ attempts; a timing-consistency check left $n{=}2958$, with exclusions and source-hash checks reported in the appendices. These results establish margins for the evaluated clusters; other endpoint agents and runtime APIs require separate timing.

\subsubsection{In-Kernel Enforcement Path}
eBPF enables in-kernel containment (\texttt{bpf\_send\_signal}): the hook issues SIGKILL inline during the adversary's syscall, so the same $\approx$9\,ms termination begins in the kernel, well below the $\approx$65\,ms capture and orders of magnitude below the $\approx$2.1\,s eviction (Fig.~\ref{fig:eval-race}(a)). The decisive difference is where enforcement occurs, not raw speed: the kill fires outside the control-plane path, so the FMP hold hook cannot sequence capture ahead of it. When signal issuance was measured separately from completed termination, userspace enforcement exhibited distinct fast and millisecond-scale modes whose mixture varied with load and run: delayed observations ($t_R{>}0.5$\,ms) were rare through $2\times$ ($\le6.2\%$ in every run), frequent and variable at $3\times$--$5\times$ ($0.8$--$85.6\%$), and rare again at $6\times$--$10\times$ ($\le0.4\%$). In-kernel enforcement held a $0.5$--$0.8$\,$\mu$s median with $0/50{,}000$ observations over $0.5$\,ms (Fig.~\ref{fig:eval-race}(b)); no userspace scheduling precedes it. The median alone masks this variation: at $4\times$ every run median sits near the fast mode while $7.5$--$46.4\%$ of observations are already delayed. The stable distinction is enforcement location: in-kernel issuance bypasses the scheduling path used by the control-plane hold.

\subsubsection{Integrated k3s Ordering Demonstration}
We deployed \sysname\ as three k3s services: a FastAPI PEP with an in-process Cedar PDP and FMP hook, a volume-backed hash-chained evidence store, and a target workload. A policy violation emits routine and targeted records, activates the containment hold, checkpoints $\approx$259\,MiB with CRIU, records the image digest, and then contains the workload. Across 30 integrated runs on the Zen~5 node, the five-step sequence (routine, targeted, hold, capture, containment) completed capture in $\approx$67\,ms while pod memory remained available for $\approx$2049\,ms during eviction: capture preceded memory loss in 30/30 runs. This establishes deployed ordering and timing margin for an FMP-controlled workload process. Integrating the separately demonstrated Docker/\texttt{runc} container-memory path into k3s and managed clusters with provider-supported checkpointing is planned. The in-kernel and self-destruction experiments above characterize containment paths that bypass the control plane.
\begin{figure}[t]
\centering
\includegraphics[width=\columnwidth]{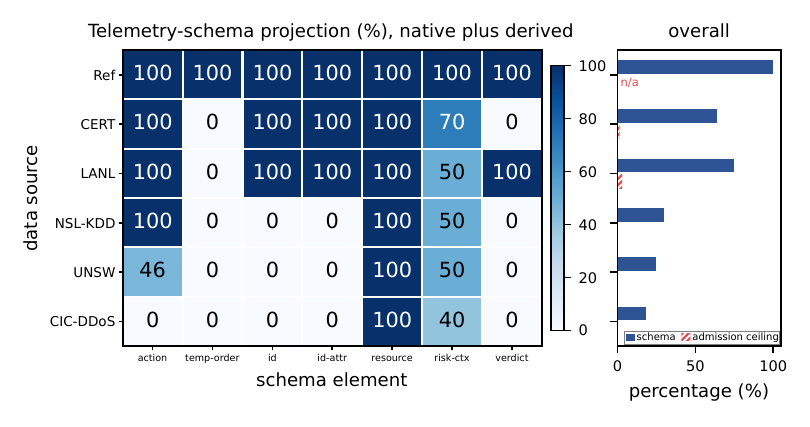}
\vspace{-8mm}
\caption{Native-plus-derived coverage for five measured sources and a synthetic schema reference: per-element projection (left); overall coverage and the separately defined admission ceiling (right; different denominators).}
\label{fig:feas-provenance}
\vspace{-5mm}
\end{figure}

\subsection{Telemetry-Schema Coverage}
\label{subsec:feas-provenance}
Capture feasibility addresses availability; telemetry-schema coverage measures how much reconstruction context the resulting records contain. Appendix~\suppref{sec:s-cost} states the replay protocol: per-source ordering, how timestamps are consumed, and why $\lambda_m$ derives from measured labels and a stipulated decision rate rather than from replay time. We replayed five labeled datasets spanning the identity--network spectrum: CERT~\cite{glasser2013cert,lindauer2020insider}, LANL~\cite{kent-2015-cyberdata1}, NSL-KDD~\cite{tavallaee2009nslkdd}, UNSW-NB15~\cite{moustafa2015unsw}, and CIC-DDoS2019~\cite{sharafaldin2019cicddos}, and scored them against a synthetic schema reference. That reference row (\emph{Ref} in Fig.~\ref{fig:feas-provenance}) is $100\%$ on every element \emph{by construction}: the ceiling a source would reach by populating every decision-record field. It is excluded from origin-basis scoring and carries no measured event count.

We measure \emph{(i) telemetry-schema coverage}, the weighted fraction of FMP decision-record elements a source can populate (identity, identity attributes, action, resource, risk context, verdict, and temporal-order resolution), crediting native and deterministic-derived fields but not imputed ones; and \emph{(ii) the full-memory admission ceiling}, the fraction of malicious events a rate-capped budget can admit. Timestamp-derived resolution is tie-aware and does not imply causal recovery (Table~\suppref{tab:s-provfields}): $r_{\mathrm{temp}}{=}(U{-}1)/(n{-}1)$ for $n$ events and $U$ distinct timestamps. Timestamp-derived ordering receives zero credit under the stated scoring rule for all five measured sources. CERT ($r_{\mathrm{temp}}{=}0.581$) and LANL ($0.00477$) are therefore reported descriptively but score zero. Among the network sources, NSL-KDD and UNSW-NB15 expose no wall-clock field, while CIC-DDoS2019 records timestamps that the evaluated adapter does not consume. Appendix~\suppref{sec:s-cost} gives the derivation, full-stream counts, and crediting rules. Identity credit requires an IdP-bound principal, not a bare IP or per-flow pseudonym.

\noindent\textbf{Metric scope.} Schema coverage measures available fields, whereas the admission ceiling measures the fraction of eligible malicious events admitted by the memory budget. They bound reconstruction context and admission rather than measure detection or end-to-end reconstruction. Tables~\suppref{tab:s-provenance} and~\suppref{tab:s-provfields} give the mappings, weights, formula, and detailed results.

\subsubsection{Context Coverage}
The FMP records action and resource for nearly every decision but obtains authenticated identities, identity attributes, and native verdict labels only when upstream telemetry supplies them. Identity-centric CERT and LANL reach 64--75\% schema coverage. Network sources using flow endpoints rather than authenticated identities reach 18--30\%. Re-scoring under four alternative weight vectors (equal, identity-heavy, risk-heavy, and one dropping temporal-order; Appendix~\suppref{sec:s-robustness}) leaves identity-rich sources at $63$--$88\%$ and network sources at $12$--$36\%$ with the ranking unchanged, so the separation persists. The missing \texttt{identity} and \texttt{identity\_attributes} fields in Fig.~\ref{fig:feas-provenance} reflect the source feed, not the FMP schema. This is the reconstruction counterpart of the identity-fragmentation challenge in Section~\ref{sec:challenges}.

\subsubsection{Full-Memory Admission}
The admission ceiling falls as eligible volume rises. The full-memory budget admits only 0.01--0.02\% of malicious events in the UNSW, NSL-KDD, and CIC-DDoS streams, which carry the highest labeled malicious fractions $h$ (all sources share the stipulated $R$), compared with 1.24\% for CERT and 2.85\% for LANL. It assumes a stationary load and perfect admission among eligible events, with no false escalations consuming budget. The cap bounds storage, while the admission policy triages eligible malicious events for full-memory capture. Section~\ref{subsec:feas-robustness} discusses severity-weighted admission.
\begin{figure}[t]
\centering
\includegraphics[width=\columnwidth]{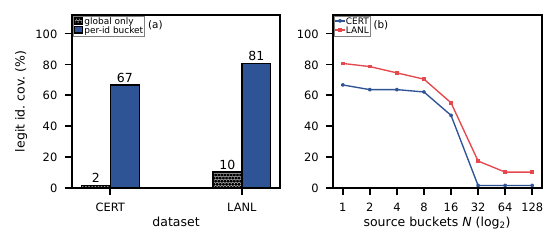}
\vspace{-8mm}
\caption{Capture-budget resilience to an A3 flood. (a) Global and per-identity budgeting; (b) legitimate-identity coverage as the flood spreads across $N$ synthetic sources.}
\label{fig:eval-ratecap}
\vspace{-4mm}
\end{figure}

\vspace{-3mm}
\subsection{Robustness and Scalability}
\label{subsec:feas-robustness}
We test self-destruction, detector quality, storage trade-offs, budget exhaustion, evidence-log scaling, and \mbox{detector-driven} recovery.

\begin{table}[t]
\caption{Trigger-decoupled recovery ($n{=}1000$/detector). \textbf{Mem.\ art.}: four planted memory-resident markers. \textbf{C2 sock.\ (1st)}: first-attempt \textsc{established}-socket recovery. \textbf{Capture}: median of each trial's terminating checkpoint, not attack-to-alert latency.}
\label{tab:detectors}
\centering\scriptsize
\setlength{\tabcolsep}{3pt}
\resizebox{\columnwidth}{!}{%
\begin{tabular}{|l|l|c|c|c|}
\hline
\textbf{Detector} & \textbf{Modality} & \textbf{Capture} & \textbf{Mem.\ art.} & \textbf{C2 sock.\ (1st)} \\
\hline
Falco (Zen~5)    & eBPF/syscall  & $\approx$0.16\,s  & 1000/1000 & 1000/1000 \\
\hline
Suricata (Zen~5) & network IDS   & $\approx$0.23\,s  & 1000/1000 & 763/1000 \\
\hline
\end{tabular}}
\vspace{-5mm}
\end{table}

\subsubsection{Adaptive Self-Destruction}
The race in Section~\ref{subsec:feas-race} initially assumes defender-driven containment, but a compromised workload may destroy its own memory after detecting scrutiny. On the Zen~5 node (Table~\suppref{tab:s-adversarial}; $n{=}1000$ per strategy), a 256\,MiB workload exited in $\approx$7\,ms median (\texttt{self\_exit}) or scrubbed memory in $\approx$3\,ms (\texttt{mem\_scrub}), versus $\approx$65\,ms reactive CRIU capture ($p95\approx75$\,ms). Both destruction distributions lay entirely below capture. A Python-level overwrite measured $\approx$67\,ms but overlapped the 55--114\,ms capture distribution, characterizing only a slower software-level scrub.

Self-destruction bypasses the orchestrator, making the $\approx$2.1\,s eviction margin irrelevant and reactive full-memory capture unreliable. The always-on routine tier remains: its record is serialized and hash-chained inline ($\approx$7\,$\mu$s), then protected after durable externalization by a signed periodic anchor. The monitored workload cannot retroactively destroy the anchored record.

\subsubsection{Detector-Quality Sensitivity}
Routine decision provenance is detector-independent. For full-memory malicious-event admission, let $r$ be detector recall and $C$ the source-specific admission ceiling. We model the admitted malicious-event fraction as $\min(r,C)$. Across the evaluated sources, the largest ceiling is $2.85\%$, so admission becomes budget-limited once recall exceeds the corresponding $C$ (Fig.~\ref{fig:feas-bounds}(b)). Let $f_{\mathrm{FP}}$ be the false-escalation fraction. False escalations consume the same budget and alone would saturate the cap when $f_{\mathrm{FP}}\ge c/R$ ($10^{-4}$ at the reference point). The appendices specify the schema-weight sensitivity check. Joint precision--recall evaluation against a live detector remains future work.

\subsubsection{Decision-Record-Only Baseline}
Relative to a decision-record-only baseline, the FMP preserves identical decision provenance while adding session artifacts and rate-capped host/memory state. These additions raise steady-state storage from $\approx$7.0\,TB to $\approx$46\,TB, a $\approx 6.6\times$ increase (Table~\suppref{tab:s-baseline}).

\subsubsection{Capture-Budget Denial of Service}
An A3 adversary pursuing TO4 can flood high-risk events to exhaust the full-memory budget. In a deterministic replay of the admission logic driven by a CIC-DDoS2019 flood interleaved with legitimate CERT and LANL incidents at the $0.05$/s cap, a global-only budget achieved legitimate-identity coverage of just $1.5\%$ (CERT) and $10.2\%$ (LANL), while the flood consumed $94$--$99\%$ of granted captures. A per-identity token bucket raised legitimate-identity coverage to $66.7\%$ and $80.6\%$ (fractions of \emph{labeled incidents} admitted in that replay, a different denominator from the all-malicious-event ceilings of Section~\ref{subsec:feas-provenance}) and cut the adversary's share to $\approx$3\% (Fig.~\ref{fig:eval-ratecap}(a)).

Per-identity budgeting is strongest against concentrated floods. As flooding spreads across $N$ synthetic sources, legitimate-identity coverage approaches the global-only baseline once $N$ nears the legitimate population (Fig.~\ref{fig:eval-ratecap}(b)). This study models source dispersion, not botnet size. The per-identity refill rate is the global budget divided by an assumed active-identity count, so fair-share allocation presumes an estimate of that population, although buckets themselves are created on first appearance rather than enumerated in advance. Production enforcement of the evaluated per-identity policy and severity-weighted admission are planned for false-positive floods.

\subsubsection{Evidence-Log Scalability} A globally locked hash chain serializes appends ($\approx$1.0\,M records/s peak, $0.6$\,M contended); per-shard chains scale near-linearly to $\approx$28\,M records/s at a flat $\approx$0.6\,$\mu$s $p99$ in an in-memory single-host benchmark that excludes networking and durable writes. In the architecture, shard state may subsequently be incorporated into a periodically signed global Merkle root~\cite{rfc9162}; the benchmark excludes this aggregation cost. Signed roots detected alteration of anchored state in 1000/1000 trials, addressing post-capture record alteration (TO3) and repudiation (TO5). This property assumes an uncompromised signing key; source falsification before chaining remains outside it. Figure~\suppref{fig:s-eval-ledger} gives the scaling curves.

\subsubsection{Detector-Driven Closed Loop and Forensic Utility}
To exercise the closed loop with two detector modalities, we drove the same \texttt{runc}/CRIU checkpoint from Falco (eBPF/syscall) and Suricata (network IDS) reacting to a fileless attack (an in-workload reverse shell to a live C2 socket plus a sensitive-file read). In the preregistered study, Falco and Suricata each triggered in 1000/1000 trials, so detection was not the limiting factor in this experiment. The first checkpoint retained the \textsc{established} C2 socket in 1000/1000 Falco trials and 763/1000 Suricata trials, and every successful checkpoint recovered all four planted memory-resident markers. Suricata's 237 first-attempt socket misses occurred despite successful checkpoints: there were no detector misses or checkpoint failures. Across repeated independent redeployment and detonation attempts, eventual recovery was 1000/1000. Driving one capture path from both modalities demonstrates trigger decoupling (Table~\ref{tab:detectors}). Falco detection reproduced 1000/1000 on managed DOKS; full-memory acquisition there depends on provider-supported checkpointing because node-level checkpointing was unavailable~\cite{kep2008}. The evaluated capture path therefore targets operator-controlled infrastructure, with managed and serverless integration planned through provider-supported checkpointing.

\subsection{Matched Comparison Against a Snapshot-Chain Baseline}
\label{subsec:feas-matched}
For a controlled comparison, we fixed the workload, alert source, containment signal, host state, and two recovery predicates defined below, then varied only acquisition policy: \textbf{Bare}, reactive acquisition with no FMP; \textbf{Unsequenced}, the FMP path with containment \emph{not} held; \textbf{Sequenced}, the FMP path with containment released at capture completion; and \textbf{Chain}, an FSC-inspired CRIU incremental-checkpoint baseline implementing the periodic policy of~\cite{stoyanov2026fsc}.
Chain is not the FSC authors' implementation: it
realizes the periodic acquisition policy only, and omits their orchestration,
filesystem and metadata capture, chain-management and integrity services, and
analysis workflow. Its absolute costs are not a measurement of their system.

Two endpoints are measured everywhere: $R_{\mathrm{mem}}$, three long-lived
markers resident from process start, and $R_{\mathrm{sock}}$, the
\textsc{established} C2 socket, which exists only once the attack opens it.
Recovery means exact post-containment extraction from the retained image or parent-resolved chain; the OCI bundle does not persist long enough for container restoration. Rates use Clopper--Pearson intervals and differences use Newcombe--Wilson hybrid-score intervals; both were fixed in advance because boundary rates invalidate Wald intervals. Because Chain yields
several snapshots per trial, each trial is reduced to one predeclared summary
before resampling; the trial, not the snapshot, is the unit of replication.

Every trial correlates its alert to its own container by exact identity and
requires the event to follow that trial's trigger; across all $1000$ trials the
minimum lead is $+0.42$\,s and every configuration is $200/200$ valid; Table~\suppref{tab:s-ad-provenance} records host, tooling, CRIU configuration and correlation discipline. Observed
start-to-start cadence is $2.000$\,s and $1.000$\,s.

\noindent\textbf{Controlled contrast.} In Table~\ref{tab:matched}, Bare and Unsequenced recovered
$0/200$ (difference $+0.000$, 95\%~CI $[-0.019,+0.019]$); Sequenced recovered
$200/200$ ($+1.000$, $[+0.973,+1.000]$). Bare invokes \texttt{runc} directly, while Unsequenced and Sequenced share the FMP's leave-running path; Bare therefore provides a \emph{descriptive} reference; Sequenced versus Unsequenced is the controlled comparison. Bare, Unsequenced, and Sequenced do not exercise the gateway policy engine
or the decision-record path, whose cost is reported in
Section~\ref{subsec:feas-results}. Bare and Unsequenced
lose because the kill arrives $\approx$3\,ms after the alert, before capture completes. Sequenced's median capture time was $0.140$\,s under the hold, so the Bare and Unsequenced losses arise from timing rather than capture-path failure.

The comparisons answer different questions. Sequenced versus Unsequenced isolates the effect of the hold within a common acquisition path; Sequenced versus Chain contrasts current event-scoped capture with prior periodic state. The first supports the ordering claim, while the second characterizes recovery and cost under two sampling policies rather than ranking complete systems.

A single-endpoint experiment at larger $n$ corroborates ordering and isolates it from slow-path advantage. When capture and the kill fire together, the planted secret is lost in every Docker-container trial (0/1000, $L{=}100\%$), while the sequencing hold of Section~\ref{subsec:sequencing} preserves it in all of them (1000/1000, $L{=}0\%$); a bare process corroborates the result (0/1000 vs.\ 1000/1000). Only sequencing preserves the artifact under the same fast kill; unsequenced acquisition, bare or FMP-mediated, does not.

The two recovery endpoints distinguish Sequenced from Chain. For long-lived evidence,
both showed $200/200$ recovery (observed difference $+0.000$, 95\%~CI
$[-0.019,+0.019]$); Chain always held a prior snapshot with the resident markers. For
the transient socket, Sequenced recovered $200/200$ and Chain $0/200$ at both
cadences. Across all $2284$ retained snapshots, none held the socket, so the outcome does not depend on how
snapshots are ordered relative to the alert. The mechanism is
\emph{when each policy samples}:
Sequenced captures after the current alert and before containment, whereas
Chain holds only
its most recent \emph{completed} periodic checkpoint, which had completed a
median $0.57$\,s before containment at 2\,s cadence and $0.45$\,s at 1\,s. These
are completed-checkpoint ages, not exact state-capture ages: the checkpoint
itself spans $\approx$0.20\,s. That sampling difference is the experimental treatment, not a measurement
artifact. The experiment therefore establishes the observed policy contrast, not a socket-lifetime or universal cadence bound: the instant the socket becomes established is not instrumented,
and Falco's event time is the alert-triggering event, not evidence appearance.

SIGKILL-to-observed-dead was under $1.6$\,ms in every condition, about two orders of
magnitude below the capture times above. Because Bare, Unsequenced, and Chain kill through a shell subprocess while Sequenced kills directly, these values characterize termination and reap costs rather than cross-policy timing or added sequencing delay.

The policies trade off containment delay, retained storage, and workload suspension. Sequenced freezes the workload once per containment
event ($121.7$\,ms), retains $9.99$\,MB for that event and nothing between
events, and adds $0.140$\,s of sequencing delay, equal to its capture time, with
durable commit measured outside the containment hold. Chain adds no sequencing delay and
retains a $9.35$\,MB root once, but thereafter accrues $59.2$\,KB/s at 2\,s
cadence and $119.7$\,KB/s at 1\,s ($3.55$ and $7.18$\,MB per 60\,s), \emph{and
freezes the workload once per cadence interval}: $2.93$\,s of suspension per
60\,s, $4.9\%$ of wall-clock, rising to $9.8\%$ at 1\,s. Neither policy dominates across all three costs; periodic checkpointing avoids containment delay but repeatedly suspends the application.

\begin{table}[t]
\caption{Matched comparison (Zen~5, $n{=}200$ attempted each, all $200/200$ valid). \textbf{Seq.\ delay} is delay \textbf{added} by holding containment. Image age is the median interval from completion of the retained checkpoint to containment; it is not exact state-capture age. Storage is per event (reactive) or a one-time root plus accrual (chain); dash: no retained image.}
\label{tab:matched}
\centering\scriptsize
\setlength{\tabcolsep}{3pt}
\resizebox{\columnwidth}{!}{%
\begin{tabular}{|l|c|c|c|c|c|}
\hline
\textbf{Condition} & \textbf{$R_{\mathrm{mem}}$} & \textbf{$R_{\mathrm{sock}}$} & \textbf{Seq.\ delay} & \textbf{Image age} & \textbf{Storage} \\
\hline
Bare (A)                & 0/200   & 0/200   & $0$\,s        & --              & --                        \\
\hline
Unsequenced (B)         & 0/200   & 0/200   & $0$\,s        & --              & --                        \\
\hline
Sequenced (C)           & 200/200 & 200/200 & $0.140$\,s    & $0$\,s by policy & $9.99$\,MB/event          \\
\hline
Chain (D), 2\,s cadence  & 200/200 & 0/200   & $0$\,s        & $0.57$\,s       & $9.35$\,MB $+\,59.2$\,KB/s \\
\hline
Chain (D), 1\,s cadence  & 200/200 & 0/200   & $0$\,s        & $0.45$\,s       & $9.35$\,MB $+\,119.7$\,KB/s \\
\hline
\end{tabular}}
\vspace{-5mm}
\end{table}

\subsection{Synthesis and Threats to Validity}
\label{subsec:feas-threats}
\rqref{rq:main} found 64--75\% schema coverage for identity sources and 18--30\% for
network sources, bounding reconstruction context. \rqref{rq:adversarial} found that reactive
capture preceded slow eviction but lost to direct or in-kernel termination and
self-destruction; sequencing captured event state, while periodic checkpoints avoided delay
but missed transient evidence. \rqref{rq:feasibility} found 1.9--3.0\% in-process throughput
cost, $\approx$46\,TB modeled storage, and full-memory admission ceilings of 0.01--2.85\%.

Five limitations define the scope of these findings:
\begin{enumerate}[\setlength{\IEEElabelindent}{0pt}\setlength{\labelsep}{0.3em}]
\setlength{\itemsep}{0pt}
\setlength{\parsep}{0pt}
\setlength{\parskip}{0pt}
\setlength{\topsep}{2pt}
  \item \emph{Measurement scope.} Gateway overhead and the primary CRIU race use synthetic workloads on individual hosts; orchestrator tests additionally span k3s and a managed cluster. CRIU captures memory and task state, but live microsegment-flow capture remains a production integration task. The primary race times a 256\,MiB workload and is corroborated on a Docker container (Section~\ref{subsec:feas-race}). Repeated-trial figures are empirical rates on a single host, not independent real-world incident samples.
  \item \emph{Storage abstraction.} Equation~(\ref{eq:storage}) estimates logical retained bytes, excluding replication, metadata, write amplification, compression, and object-store costs.
  \item \emph{Baseline fidelity.} Chain reproduces FSC's periodic incremental-memory policy, not its code or complete system; the comparison therefore concerns sampling policy rather than absolute system cost.
  \item \emph{Detector dependence.} Overall recall remains bounded by upstream detection. The recall and false-escalation sensitivity analyses (Section~\ref{subsec:feas-robustness}) are analytical projections rather than a joint precision--recall study.
  \item \emph{Trust and failure semantics.} The prototype trusts the FMP capture path and co-locates its hook with the PEP. It canonically serializes and hash-chains the routine record outside the monitored workload before capture begins, so workload self-destruction cannot remove that record. Gateway or node compromise can suppress it, and failure before queued export can lose pending records; the measured $\approx$7\,$\mu$s increment excludes synchronous commit. The production architecture of Section~\ref{sec:fmp-arch} separates the monitored environment from the FMP and evidence root and specifies local capture-window expiry to bound held delay. Future work will evaluate coordinator and evidence-path failure, release semantics, recovery idempotence, and the cost of stronger commit policies (\mbox{Supplement~\suppref{sec:s-cost}}).
\end{enumerate}

\section{Conclusion}
\label{sec:conclusion}
ZT changes not only what evidence exists but whether volatile evidence survives long enough to be captured. \sysname\ makes tiered, identity- and policy-linked preservation a ZT control-plane function through an FMP that records each decision and sequences richer capture ahead of defender-routed destructive containment. The capture--containment race defines the boundary: the routine record is the minimum surviving evidence, while reactive volatile-state capture succeeds only before defender or adversary action destroys the target state.

Sequencing determined the outcome on the evaluated paths. Direct SIGKILL outran concurrent capture in all 1000 trials; sequencing preserved the artifact by completing capture before releasing that kill in all 1000. Within the common FMP acquisition path, Unsequenced recovered $0/200$ and Sequenced $200/200$. The FSC-inspired chain showed the same observed recovery on long-lived resident evidence ($200/200$) but missed the transient artifact at both cadences because it retained a prior periodic snapshot rather than state captured at the containment decision. These policies expose a three-way tradeoff: event-scoped sequencing adds containment delay and per-incident storage, while periodic capture avoids that delay but repeatedly suspends the workload and accumulates retained state.

Operationally, the always-on decision record reduced in-process throughput by 1.9--3.0\%, while replayed identity-oriented sources populated 64--75\% of its schema and rate limiting bounded full-memory admission.

The evaluated in-kernel enforcement path and adversarial self-destruction bypass the sequencing barrier. \sysname\ establishes capture before defender-routed containment as a measurable forensic-readiness requirement.

\appendices
\noindent These appendices contain additional methodological detail, tables, and figures referenced by the main text.


\begin{figure*}[!h]
\centering
\resizebox{0.98\textwidth}{!}{%
\begin{tikzpicture}[
  header/.style={draw=black!70, rounded corners=2pt,
                minimum height=4mm, align=center, font=\footnotesize\bfseries,
                inner xsep=1.5pt, inner ysep=1pt, fill=black!8},
  control/.style={draw=blue!55!black, rounded corners=2pt, text width=4.7cm,
                 minimum height=7mm, align=left, font=\scriptsize,
                 inner xsep=1.5pt, inner ysep=1pt, fill=blue!5},
  loss/.style={draw=red!60!black, rounded corners=2pt, text width=3.9cm,
              minimum height=7mm, align=left, font=\scriptsize,
              inner xsep=1.5pt, inner ysep=1pt, fill=red!4},
  mitigation/.style={draw=green!45!black, rounded corners=2pt, text width=3.9cm,
                    minimum height=7mm, align=left, font=\scriptsize,
                    inner xsep=1.5pt, inner ysep=1pt, fill=green!5},
  maparrow/.style={-{Stealth[length=2mm]}, thick, black!55}
]
  \node[header,text width=4.7cm] (hc) at (0,0) {ZT control};
  \node[header,text width=3.9cm] (hl) at (5.1,0) {Evidence limitation};
  \node[header,text width=3.9cm] (hm) at (9.8,0) {FMP response};

  \node[control] (c1) at (0,-0.85)
    {\textbf{Continuous verification} Short-lived tokens; changing attributes};
  \node[loss] (l1) at (5.1,-0.85)
    {Identity and policy data spans systems and may not be retained.};
  \node[mitigation] (m1) at (9.8,-0.85)
    {Persist identity, action, resource, policy, trace, and verdict.};

  \node[control] (c2) at (0,-1.80)
    {\textbf{Microsegmentation and mTLS} Distributed PEPs; encrypted traffic};
  \node[loss] (l2) at (5.1,-1.80)
    {Flow evidence is split across PEPs; payloads remain encrypted.};
  \node[mitigation] (m2) at (9.8,-1.80)
    {Correlate identity and flow context; trigger targeted capture.};

  \node[control] (c3) at (0,-2.75)
    {\textbf{Ephemeral workloads} Automated restart, eviction, or termination};
  \node[loss] (l3) at (5.1,-2.75)
    {Memory, writable layers, and flows may disappear before capture.};
  \node[mitigation] (m3) at (9.8,-2.75)
    {Gate destructive containment on bounded, rate-limited capture.};

  \node[control] (c4) at (0,-3.70)
    {\textbf{Least privilege and JIT} Temporary access; protected evidence stores};
  \node[loss] (l4) at (5.1,-3.70)
    {Access expires; direct analysis of master artifacts risks alteration.};
  \node[mitigation] (m4) at (9.8,-3.70)
    {Broker JIT access to verified copies; retain a sealed, append-only master.};

  \foreach \n in {1,2,3,4}{
    \draw[maparrow] (c\n.east) -- (l\n.west);
    \draw[maparrow] (l\n.east) -- (m\n.west);
  }
\end{tikzpicture}%
}
\captionsetup{font=footnotesize,skip=3pt}
\caption{Representative ZT controls, associated evidence limitations, and FMP responses.}
\label{fig:s-forensic-gap}
\end{figure*}
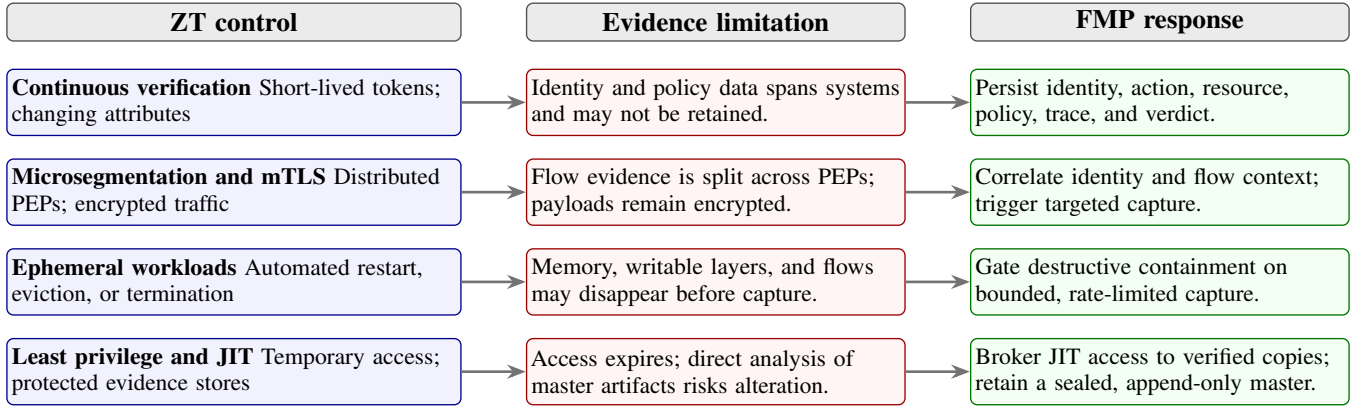

\section{Evidentiary and Legal Mapping}
\label{sec:s-legal}
The FMP is designed to support authentication and integrity showings under the Federal Rules of Evidence (FRE)~\cite{fre} and analogous Military Rules of Evidence~\cite{mcm2024}, and preservation and discovery under the Federal Rules of Civil Procedure (FRCP)~\cite{frcp}. ISO/IEC~27037 and 27041--27043 provide related guidance~\cite{iso27037,iso27041,iso27042,iso27043}; remote-forensics work likewise addresses chain of custody and admissibility~\cite{chew2024lawful}. Tables~\ref{tab:s-legal-provisions} and~\ref{tab:s-evidence-mapping} summarize the authorities and map them to FMP capabilities. \begin{table*}[t]
\captionsetup{font=footnotesize,skip=3pt}
\caption{Selected federal evidentiary and civil-procedure authorities relevant to the architecture; Table~\ref{tab:s-evidence-mapping} also cites FRCP~26 and~34 for discovery scope and electronically stored information (ESI) production.}
\label{tab:s-legal-provisions}
\centering
\footnotesize
\renewcommand{\arraystretch}{0.8}
\begin{tabularx}{\textwidth}{|l|X|}
\hline
\textbf{Rule or authority} & \textbf{Substance} \\
\hline
FRE 901(a) & General authentication: the proponent must produce evidence sufficient to support a finding that the item is what it is claimed to be. \\
\hline
FRE 901(b)(9) & Authentication by ``evidence describing a process or system and showing that it produces an accurate result.'' \\
\hline
FRE 902(13) & Electronic-process records may be self-authenticated through certification by a qualified person (added 2017). \\
\hline
FRE 902(14) & Copied electronic data may be self-authenticated through digital identification and certification by a qualified person (added 2017). \\
\hline
FRE 803(6) & Business-records exception, subject to timing, knowledge, regular course and practice, foundation, and trustworthiness. \\
\hline
FRE 1001(d) & For ESI, an ``original'' includes a sight-readable output that accurately reflects the information. \\
\hline
FRE 1003 & Duplicates are admissible as originals unless authenticity is genuinely disputed or admission would be unfair. \\
\hline
FRE 702 \& \emph{Daubert} & Expert-testimony reliability: nonexclusive factors include testability, peer review, error rate and standards, and general acceptance~\cite{daubert1993}. \\
\hline
FRCP 37(e) & Applies when ESI that should have been preserved is lost because reasonable steps were not taken and it cannot be restored or replaced: curative measures require prejudice; listed severe measures require intent to deprive. \\
\hline
\end{tabularx}
\end{table*}

\begin{table*}[t]
\captionsetup{font=footnotesize,skip=3pt}
\caption{FMP capability mapping to evidentiary authorities and technical guidance.}
\label{tab:s-evidence-mapping}
\centering
\scriptsize
\renewcommand{\arraystretch}{1.0}
\begin{tabularx}{\textwidth}{|>{\raggedright\arraybackslash}p{0.2\textwidth}|>{\raggedright\arraybackslash}p{0.27\textwidth}|>{\raggedright\arraybackslash}X|}
\hline
\textbf{FMP capability} & \textbf{Relevant authority/guidance} & \textbf{Technical support} \\
\hline
\multirow{1}{=}{Policy-triggered state capture} & FRE 803(6), 902(13); FRCP 37(e); ISO/IEC 27037, 27043 & Automated records supporting a regular-practice foundation, Rule~902(13) certification, and reasonable preservation steps. \\
\hline
Identity- and policy-linked reconstruction & FRE 901(a), 902(13);\newline ISO/IEC 27042 & Identity, action, and policy-decision records supporting attribution and reconstruction. \\
\hline
Automated telemetry orchestration & FRE 803(6), 901(b)(9); FRCP 26, 34; ISO/IEC 27041, 27043 & A documented, reproducible collection process that assembles consistent and producible records. \\
\hline
\multirow{1}{=}{JIT access to working copies} & FRE 1001(d), 1003;\newline ISO/IEC 27037 & A sealed master and verified duplicates that protect the original evidence from modification. \\
\hline
Hash-linked records and signed anchors & FRE 901(a), 902(13), 902(14);\newline ISO/IEC 27037, 27042 & Hashes, signatures, and handling records supporting authentication of system-generated records and copied artifacts. \\
\hline
FMP administration and maintenance & FRE 901(b)(9), 803(6); FRCP 26, 34; ISO/IEC 27037, 27041 & Policy-as-code, brokered administration, and data-class tags supporting process reliability, retention, and redaction. \\
\hline
\end{tabularx}
\end{table*}

\noindent FMP records and controls may support authentication, preservation, and process showings; courts decide admissibility. Rules~902(13)/(14) address authentication only and require certification by a qualified person and Rule~902(11) notice; hashes and signatures alone do not establish admissibility.



\section{Documented Identity and Gateway Evidence Gaps}
\label{sec:s-scenarios}
Fig.~\ref{fig:s-forensic-gap} links representative ZT controls to evidence limitations and FMP responses. Table~\ref{tab:s-incidents} relates identity-misuse and gateway-compromise incidents to their investigative implications. The main paper analyzes encryption in Section~\mainref{sec:challenges} and evaluates ephemerality in Section~\mainref{sec:feasibility}.

The cited reports and advisories document incident mechanisms and records available to investigators. They motivate the design; the main paper evaluates it. Valid-session abuse and compromised gateways can leave reconstruction dependent on incomplete or dispersed records. \sysname\ therefore records each policy decision and can trigger preservation before destructive response. \begin{table*}[t]
\captionsetup{font=footnotesize,skip=3pt}
\caption{Documented incidents and their identity- or gateway-evidence implications. MFA is multi-factor authentication; SSO is single sign-on.}
\label{tab:s-incidents}
\centering
\footnotesize
\setlength{\tabcolsep}{3pt}
\renewcommand{\arraystretch}{0.88}
\begin{tabularx}{\textwidth}{|>{\raggedright\arraybackslash}p{0.15\textwidth}|c|>{\raggedright\arraybackslash}p{0.26\textwidth}|>{\raggedright\arraybackslash}X|>{\raggedright\arraybackslash}p{0.06\textwidth}|}
\hline
\textbf{Incident} & \textbf{Year} & \textbf{Mechanism abused} & \textbf{Investigative implication} & \textbf{Source} \\
\hline
\multicolumn{5}{|l|}{\emph{Identity misuse: valid tokens or credentials enable apparently authorized activity}} \\
\hline
SolarWinds / Golden SAML & 2020 & Forged SAML assertions from a stolen token-signing key & Forged tokens bypassed passwords and MFA; detection required correlating cloud sign-ins with AD FS and domain-controller events & \cite{cisa2021aa21008a,mandiant2021unc2452} \\
\hline
Okta support breach & 2023 & Post-MFA session cookies replayed from stolen HTTP Archive (HAR) files & Replay impersonated valid users; hunting used session, user, IP, and System Log data & \cite{okta2023har} \\
\hline
Midnight Blizzard & 2024 & Malicious OAuth applications granted \texttt{full\_access\_as\_app} & OAuth-authorized Exchange Web Services (EWS) activity required audit-log analysis; changing proxy IPs weakened conventional IOC searches & \cite{msft2024midnightblizzard} \\
\hline
Snowflake / UNC5537 & 2024 & Stolen credentials used against tenants without MFA & Stolen credentials and native clients made authentication, client, session, and query logs central to reconstruction & \cite{mandiant2024snowflake} \\
\hline
CircleCI SSO-cookie theft & 2023 & Malware exfiltrated a valid, MFA-authenticated SSO session cookie & Impersonation via a valid session; the investigation correlated authentication, network, monitoring, and system-log records & \cite{circleci2023incident} \\
\hline
\multicolumn{5}{|l|}{\emph{Gateway compromise: session replay and attacker-altered appliance evidence}} \\
\hline
Citrix Bleed & 2023 & NetScaler session tokens stolen from memory and replayed to bypass MFA & Replayed cookies authenticated without username, password, or MFA tokens; hunting used session and network anomalies & \cite{cisa2023aa23325a} \\
\hline
Ivanti Connect Secure & 2024 & Gateway authentication bypass and command injection & Actors deleted logs and core dumps and could deceive integrity checks, reducing confidence in appliance-local evidence & \cite{cisa2024aa24060b,mandiant2024ivanti} \\
\hline
\end{tabularx}
\vspace{-4mm}
\end{table*}


\section{Detailed Cost, Storage, and Race Parameters}
\label{sec:s-cost}
This section provides the rates, sizes, retentions, and timings behind the Section~\mainref{sec:feasibility} feasibility results.

\paragraph{Tier parameters}
\begin{table}[t]
\centering
\captionsetup{font=footnotesize,skip=3pt}
\caption{Reference rates, sizes, and retentions at $R{=}500$ decisions/s, with associated measured operation latencies ($p50$).}
\label{tab:s-tier-params}
\scriptsize
\setlength{\tabcolsep}{2.5pt}
\begin{tabular}{|l|r|r|r|r|r|}
\hline
\textbf{Tier} & \textbf{Rate (/s)} & \textbf{Bytes/ev.} & \textbf{Latency} & \textbf{Retention} & \textbf{Footprint} \\
\hline
Routine (inline)    & 500   & 1.5\,KiB & 7.1\,$\mu$s & 365\,d & 24\,TB \\
\hline
Targeted (async)    & 71.95 & 64\,KiB  & 0.08\,ms    & 90\,d  & 37\,TB \\
\hline
Full-memory (async) & 0.05  & 256\,MiB & 65\,ms      & 30\,d  & 35\,TB \\
\hline
\multicolumn{5}{|l|}{Total steady-state footprint} & 95.7\,TB \\
\hline
\end{tabular}
\end{table}
Table~\ref{tab:s-tier-params} uses conservative routine and targeted sizes, a 256\,MiB full-memory reference, and measured latencies. \emph{Full-memory} denotes the CRIU memory/task checkpoint; \emph{architectural full-state} also includes filesystem and flow artifacts. The 71.95/s targeted rate follows main Eq.~\mainref{eq:spill}. At the stated rates and 365/90/30-day retentions, Eq.~\mainref{eq:storage} yields 95.7\,TB. Measured sizes are $\approx$445\,B routine, $\approx$7.2\,KiB targeted, and $\approx$259\,MiB full-memory, yielding $\approx$46.3\,TB at the same rates and retentions. Sizes use binary units; aggregate footprints use decimal TB.

\paragraph{Per-source schema projection}
\begin{table}[t]
\centering
\captionsetup{font=footnotesize,skip=3pt}
\caption{Per-source direct-or-derived schema projection and the rate-cap-limited admission ceiling. The synthetic row defines schema completeness and is not an empirical incident stream, so its admission ceiling is n/a.}
\label{tab:s-provenance}
\setlength{\tabcolsep}{1.5pt}
\scriptsize
\begin{tabular}{|l|l|r|r|}
\hline
\textbf{Source} & \textbf{Type} & \textbf{Schema cov.\ (\%)} & \textbf{Adm.\ ceil.\ (\%)} \\
\hline
Synthetic & ref. & 100.0 & n/a \\
\hline
LANL      & identity & 75.0 & 2.85 \\
\hline
CERT      & identity & 64.0 & 1.24 \\
\hline
NSL-KDD   & network  & 30.0 & 0.02 \\
\hline
UNSW-NB15 & network  & 24.6 & 0.01 \\
\hline
CIC-DDoS  & network  & 18.0 & 0.02 \\
\hline
\end{tabular}
\end{table}
Table~\ref{tab:s-provenance} gives the per-source values summarized in Section~\mainref{subsec:feas-provenance}. The admission ceiling is $\min(1,c/\lambda_m)$ for full-capture budget $c{=}0.05$/s and per-source malicious-event rate $\lambda_m$. Across the replays, $\lambda_m$ ranges from $\approx$1.8/s (LANL) to $\approx$340/s (UNSW). This roughly 190-fold difference in $\lambda_m$, not detection, produces the source-specific admission ceilings.

\paragraph{Live-gateway records} The values above project archival datasets onto the decision record; low coverage reflects their retained fields, not the record schema. On the evaluated live gateway, \texttt{gateway\_source.py} natively populates identity, identity attributes, action, resource, risk context, and verdict from the request and policy response. These six fields carry $0.85$ of the reference weight; recorder timestamp granularity determines temporal-order resolution. The separate failure-semantics test records only case identifier, decision, request identifier, and recording timestamp and does not produce complete decision records.


\paragraph{Barrier and release-policy protocol}
All barrier runs are same-host Docker/\texttt{runc}/CRIU, not Kubernetes and not a remote production evidence store.
\emph{Synchronization point.} At $t{=}0$, each condition generates capture and containment requests. The no-barrier condition delivers direct-PID \texttt{SIGKILL} immediately; barrier conditions delay delivery to the specified release boundary.
\emph{Adversary operation.} The in-container workload runs a bounded local-write loop; one operation is a single
64\,KiB \texttt{pwrite()} recorded in a shared counter read immediately before and after the hold. It is deliberately a
local write, not network exfiltration, and bounds attacker \emph{throughput} during the hold rather than any specific
exfiltration semantics.
\emph{Equal-delay control.} A \texttt{freerun} control condition runs the workload ungated for 300\,ms to measure its
uncontained operation rate. Expected operations for any other condition are that rate multiplied by its median hold, and
activity reduction is $100(1-\text{measured}/\text{expected})$, so the comparison is against an equal delay rather than
against no delay.
\emph{Release boundaries.} \texttt{barrier} releases at checkpoint completion; \texttt{barrier\_durable} releases after
durable-store acknowledgment with the workload resumed during the flush; \texttt{barrier\_frozen\_durable} releases
after the same acknowledgment while the container cgroup stays paused.
\emph{Success criterion.} The evidence token is injected through a FIFO after startup, never via argv or environment,
and recovery scans only CRIU \texttt{pages-*.img} files. A repetition counts as success only if the checkpoint
succeeded and the token is found; for the durable conditions it must be present in \emph{both} the source and the committed
copies. The token search runs after release and never delays it.
\emph{Sample size.} $n{=}1000$ per condition; every attempted repetition, including failures and timeouts, is retained.

\paragraph{Storage and admission sensitivity}
Table~\ref{tab:s-sens} varies one governing input per row around the reference point, holding the others fixed. Each tier is linear in artifact rate, size, and retention; $R$ changes only routine and targeted storage. At the reference point, the full-memory tier contributes $34.8$\,TB; below $c\approx10^{-2}$/s, the total approaches its non-full-state minimum. With burst multiplier $\beta$ and false-escalation fraction $f_{\mathrm{FP}}$, the admission ceiling is $\min(1,c/(\beta hR+f_{\mathrm{FP}}R))$; false escalations alone saturate the cap when $f_{\mathrm{FP}}\ge c/R{=}10^{-4}$. At $\beta{=}1$, 5, and 20, the ceiling is $0.5\%$, $0.1\%$, and $0.025\%$, respectively.

\begin{table}[h]
\captionsetup{font=footnotesize,skip=3pt}
\caption{Storage and admission sensitivity around the reference point ($R{=}500$/s, $h{=}0.02$, $f_t{=}0.124$,
$c{=}0.05$/s, 256\,MiB, 365/90/30\,d), which yields 95.7\,TB. One parameter varies per row.}
\label{tab:s-sens}
\centering\scriptsize
\setlength{\tabcolsep}{4pt}
\resizebox{\columnwidth}{!}{%
\begin{tabular}{|l|r|r|r|}
\hline
\textbf{Parameter} & \textbf{low} & \textbf{reference} & \textbf{high} \\
\hline
full-memory artifact (MiB) & 128 / 78.3\,TB & 256 / 95.7\,TB & 8192 / 1174\,TB \\
\hline
full-memory rate cap (/s)        & 0.005 / 64.4\,TB & 0.05 / 95.7\,TB & 1.0 / 756\,TB \\
\hline
full-memory retention (d)   & 7 / 69.0\,TB & 30 / 95.7\,TB & 365 / 484\,TB \\
\hline
decision rate $R$ (/s)     & 100 / 46.9\,TB & 500 / 95.7\,TB & 5000 / 644\,TB \\
\hline
high-risk fraction $h$     & 0.005 / 91.9\,TB & 0.02 / 95.7\,TB & 0.1 / 116\,TB \\
\hline
targeted fraction $f_t$    & 0.05 / 76.8\,TB & 0.124 / 95.7\,TB & 0.5 / 191\,TB \\
\hline
\end{tabular}}
\end{table}

\paragraph{Reference-parameter provenance}
The reference point mixes measured and stipulated values. \emph{Measured:} 445\,B routine, 7.2\,KiB targeted, 259\,MiB full-memory, and all reported latencies. \emph{Reference sizes:} 1.5\,KiB, 64\,KiB, and 256\,MiB yield 95.7\,TB; measured sizes yield 46.3\,TB. \emph{Operator-selected:} $c{=}0.05$/s and 365/90/30-day retentions. \emph{Illustrative:} $R{=}500$/s, $h{=}0.02$, and $f_t{=}0.124$.

\paragraph{Capture-method scaling coefficients}
Acquisition medians are well described over 128/256/512/1024\,MiB ($n{=}200$/method/size/host) by $T{=}a+bS$, where $a$ is a fitted intercept and $b$ a size coefficient; these are descriptive coefficients, not isolated setup or hardware-bandwidth measurements because method invocation differs. Zen~5: CRIU $a{=}35.9$\,ms, $b{=}0.1692$\,ms/MiB (implied $5909$\,MiB/s); \texttt{process\_vm\_readv} $14.4$, $0.2143$ ($4666$\,MiB/s); \texttt{gcore} $68.7$, $0.2302$ ($4345$\,MiB/s). EPYC~7702/A100: CRIU $45.4$, $0.2907$ ($3440$\,MiB/s); \texttt{process\_vm\_readv} $25.6$, $0.4404$ ($2271$\,MiB/s); \texttt{gcore} $116.5$, $0.4991$ ($2004$\,MiB/s), with $R^2\ge0.997$. The fitted CRIU/\texttt{process\_vm\_readv} crossover is $\approx$477\,MiB on Zen~5 and $\approx$132\,MiB on the older host; the independent 512\,MiB Zen~5 schedules give near-zero gaps ($+1.7$ and $-0.3$\,ms), consistent with the first prediction.

\paragraph{Planned dependability evaluation}
Future work will evaluate four classes: \emph{coordinator failure} (an FMP crash while containment is held), \emph{evidence-path failure} (an unavailable store, saturated queue, or FMP--store partition), \emph{release semantics} (fail-open versus fail-closed when capture cannot complete), and \emph{recovery idempotence} (stale or replayed operations after restart). The capture window already limits held delay under release failure.

\begin{table}[t]
\centering
\captionsetup{font=footnotesize,skip=3pt}
\caption{Per-element completeness $c_e$ (\%) and weights $w$ under the native-plus-derived basis; the bottom row is $\mathrm{cov}{=}\sum_e w_e c_e$.}
\label{tab:s-provfields}
\scriptsize
\resizebox{\columnwidth}{!}{%
\begin{tabular}{|l|c|c|c|c|c|c|c|}
\hline
\textbf{Element} & \textbf{$w$} & \textbf{Syn} & \textbf{CERT} & \textbf{LANL} & \textbf{NSL} & \textbf{UNSW} & \textbf{CIC} \\
\hline
identity            & 0.15 & 100 & 100 & 100 & 0   & 0   & 0   \\
\hline
identity\_attr.\    & 0.15 & 100 & 100 & 100 & 0   & 0   & 0   \\
\hline
action              & 0.10 & 100 & 100 & 100 & 100 & 46  & 0   \\
\hline
resource            & 0.10 & 100 & 100 & 100 & 100 & 100 & 100 \\
\hline
risk\_context       & 0.20 & 100 & 70  & 50  & 50  & 50  & 40  \\
\hline
verdict             & 0.15 & 100 & 0   & 100 & 0   & 0   & 0   \\
\hline
temporal\_order     & 0.15 & 100 & 0   & 0   & 0   & 0   & 0   \\
\hline
\textbf{cov} (\%)   & 1.00 & \textbf{100.0} & \textbf{64.0} & \textbf{75.0} & \textbf{30.0} & \textbf{24.6} & \textbf{18.0} \\
\hline
\end{tabular}%
}
\end{table}

\begin{table}[!t]
\captionsetup{font=footnotesize,skip=3pt}
\caption{Adversary--objective--goal mapping.}
\label{tab:s-threat-summary}
\centering
\footnotesize
\setlength{\tabcolsep}{2.5pt}
\begin{tabularx}{\columnwidth}{|>{\centering\arraybackslash}p{0.65cm}|X|>{\centering\arraybackslash}p{1.25cm}|>{\centering\arraybackslash}p{1.15cm}|}
\hline
\multicolumn{1}{|c|}{\textbf{A\#}} & \multicolumn{1}{c|}{\textbf{Access and behavior}} & \multicolumn{1}{c|}{\textbf{TO\#}} & \multicolumn{1}{c|}{\textbf{G\#}} \\
\hline
A1 & Stolen credential or compromised host & 1--3, 5 & 1--4 \\
\hline
A2 & Malicious insider & 2, 3, 5 & 2--4 \\
\hline
A3 & Anti-forensic capability exercised by A1/A2 & 1--5 & 1--5 \\
\hline
\end{tabularx}
\end{table}

\paragraph{Temporal-order scoring}
The temporal-order element of Section~\mainref{subsec:feas-provenance} measures timestamp resolution, not causal recovery. For event timestamp $s_i$ let $c_{s_i}$ be its tie-bucket size, $n$ the event count, and $U$ the number of distinct timestamps. Local per-event resolution is $n^{-1}\sum_i c_{s_i}^{-1}{=}U/n$; we report the endpoint-normalized $r_{\mathrm{temp}}{=}(U{-}1)/(n{-}1)$ for $n>1$ and 0 otherwise, so a source with all-unique timestamps scores 1 and one fully tied bucket scores 0. Because fixed-size replay subsampling thins timestamp collisions, the temporal metric is computed separately at native event density over the full identity-source streams: CERT has $n{=}1{,}260{,}239$, $U{=}732{,}011$, and $r_{\mathrm{temp}}{=}0.580851$; LANL has $n{=}1{,}051{,}430{,}459$, $U{=}5{,}011{,}198$, and $r_{\mathrm{temp}}{=}0.004766$. All five sources receive zero temporal-order credit under every scoring basis. CERT and LANL retain the descriptive resolutions above; CIC-DDoS2019 carries a timestamp field the evaluated adapter does not consume, while NSL-KDD and UNSW-NB15 contain no wall-clock timestamp. These full-stream values therefore do not change any coverage score. The remaining elements are credited as follows: identity requires an IdP-bound principal, not a bare IP or per-flow pseudonym; risk context is the per-event fraction of available risk signals; and action, resource, and verdict each require a non-empty projected field.

\paragraph{Barrier vs.\ no barrier across size and host}
Across workload sizes, the barrier preserves the token and the unheld condition loses it. On the second host, release-policy ordering is unchanged; only absolute holds scale.

\begin{table}[t]
\captionsetup{font=footnotesize,skip=3pt}
\caption{Barrier vs.\ no barrier across workload size ($n{=}100$ per condition and size; \texttt{runc}/CRIU checkpoint, direct-PID SIGKILL, token injected after startup and recovered only from \texttt{pages-*.img}). $L$ is planted-secret loss.}
\label{tab:s-sizesweep}
\centering\footnotesize
\setlength{\tabcolsep}{6pt}
\begin{tabular}{|r|c|c|}
\hline
\textbf{RSS} & \textbf{no-barrier $L$} & \textbf{barrier $L$} \\
\hline
256\,MiB & 100\% & 0\% \\
\hline
512\,MiB & 100\% & 0\% \\
\hline
1\,GiB & 100\% & 0\% \\
\hline
2\,GiB & 100\% & 0\% \\
\hline
4\,GiB & 100\% & 0\% \\
\hline
8\,GiB & 100\% & 0\% \\
\hline
\end{tabular}
\end{table}

\begin{table}[t]
\captionsetup{font=footnotesize,skip=3pt}
\caption{Cross-host container barrier experiment on the EPYC~7702/A100 host ($n{=}100$ per condition). Holds are medians; activity reduction is versus an equal ungated delay; $L$ is planted-secret loss.}
\label{tab:s-crosshost}
\centering\footnotesize
\setlength{\tabcolsep}{4pt}
\begin{tabular}{|l|l|r|r|c|}
\hline
\textbf{RSS} & \textbf{Release point} & \textbf{Hold} & \textbf{Act.\ red.} & \textbf{$L$} \\
\hline
256\,MiB & checkpoint       & 203\,ms  & 94.7\% & 0\% \\
\hline
256\,MiB & durable-resume   & 756\,ms  & 25.9\% & 0\% \\
\hline
256\,MiB & frozen-durable   & 786\,ms  & 96.5\% & 0\% \\
\hline
1\,GiB   & checkpoint       & 437\,ms  & 96.6\% & 0\% \\
\hline
1\,GiB   & durable-resume   & 2372\,ms & 18.6\% & 0\% \\
\hline
1\,GiB   & frozen-durable   & 2408\,ms & 98.5\% & 0\% \\
\hline
\end{tabular}
\end{table}

\section{Regulated Data Handling}
\label{sec:s-regulated}
The following controls define production requirements. For evidence containing electronic protected health information (ePHI), \sysname\ applies encryption, access control, and decryption logging. HIPAA's six-year period applies to Security Rule documentation; artifact retention follows applicable policy~\cite{hipaa_security}. For PCI DSS account data, \sysname\ minimizes retention, renders stored primary account numbers (PANs) unreadable, and masks displayed PANs. When used to reduce assessment scope, segmentation requires validation~\cite{pcidss}. Data-class labels select retention, masking, and access policies.

\FloatBarrier

\begin{table*}[h]
\captionsetup{font=footnotesize,skip=3pt}
\caption{Threats, FMP mechanisms, and supporting evidence (main-paper threat model, \S II).}
\label{tab:s-threat-map}
\centering
\footnotesize
\renewcommand{\arraystretch}{0.88}
\begin{tabularx}{\textwidth}{|l|X|l|}
\hline
\textbf{Threat / goal} & \textbf{FMP mechanism} & \textbf{Evidence / analysis} \\
\hline
TO1 evidence destruction (race) & Capture barrier gates containment on capture completion & Main~\S\S~VI-C, VI-D \\
\hline
TO2 evasion of richer capture & Always-on decision record; signal-triggered targeted preservation & Main~\S\S~VI-A, VI-E \\
\hline
TO3 post-capture record alteration & Hash-linked records plus periodically signed anchors & Main~\S VI-F \\
\hline
TO4 capture DoS & Global rate cap plus per-identity admission budgeting & Main~\S VI-F \\
\hline
TO5 repudiation & Available identity context plus signed anchors & Main~\S VI-F \\
\hline
Residual: incomplete capture & Rate-capped full-memory admission bounds availability & Main~\S VI-E \\
\hline
\end{tabularx}
\end{table*}

\begin{table*}[h]
\captionsetup{font=footnotesize,skip=3pt}
\caption{Evaluation hosts. DOKS is DigitalOcean Kubernetes Service.}
\label{tab:s-nodes}
\centering
\scriptsize
\setlength{\tabcolsep}{4pt}
\renewcommand{\arraystretch}{0.9}
\begin{tabularx}{\textwidth}{|l|X|X|X|}
\hline
\textbf{Node} & \textbf{Hardware} & \textbf{Kernel / runtime} & \textbf{Used for} \\
\hline
Zen~5 & AMD Ryzen Threadripper 9960X (24C/48T), 125\,GB & Ubuntu 24.04.4; kernel 7.0.0-28, 7.0.0-29 for the A--D study and 6.17.0-40 for microbench/gateway runs; runc~1.3.4, CRIU~4.2 & Capture, containment, eBPF, adversary, evidence log, overhead, integrated k3s \\
\hline
GPU server & AMD EPYC 7702 (16\,vCPU) + NVIDIA A100 80\,GB (drv~595.71.05), 125\,GB & Ubuntu 24.04.4, 6.17.0-35; runc~1.3.4, CRIU~4.2, Docker~29.1.3 & Dataset replay (coverage), behavioral end-to-end, capture cross-check \\
\hline
DOKS & $3{\times}$ s-2vcpu-4gb (2\,vCPU / 4\,GB / 80\,GB SSD), NYC1 & Debian~13, 6.12.96; K8s~1.36.3-do.0, containerd~2.2.3 & Managed eviction, detector leg \\
\hline
\end{tabularx}
\end{table*}

\begin{table*}[h]
\captionsetup{font=footnotesize,skip=3pt}
\caption{Per-experiment tooling, replication, and analytic scope.}
\label{tab:s-testbed}
\centering
\scriptsize
\setlength{\tabcolsep}{4pt}
\renewcommand{\arraystretch}{0.8}
\begin{tabular}{|l|l|r|}
\hline
\textbf{Experiment} & \textbf{Measured tooling} & \textbf{$n$ / scope} \\
\hline
Capture-primitive microbench + cost composition & serialization; OpenSSL~3.0.13 & $3{\times}10^2$--$5{\times}10^4$ \\
\hline
Live-gateway overhead & OPA, Cedar, Casbin & 30 each \\
\hline
Concurrency study (32--256) & Cedar, EPYC host & 20 each \\
\hline
Containment race (SIGKILL vs graceful) & POSIX signals & 1000/mode \\
\hline
Full-memory CRIU checkpoint & CRIU~4.2, \texttt{runc}~1.3.4 & 1000 \\
\hline
Matched-boundary capture methods & CRIU / \texttt{gcore} / \texttt{process\_vm\_readv} & 200/method $\times$ 4 sizes $\times$ 2 hosts \\
\hline
Adaptive self-destruction & exit / \texttt{memset} / Python & 1000/strategy \\
\hline
Kubernetes eviction & k3s~v1.36.2+k3s1 & 100 forced, 100 graceful \\
\hline
Capture-vs-containment race sampler & CRIU~4.2 + SIGKILL/evict & 1000 per range \\
\hline
Remediation profiles (container stop; revocation) & Docker; control plane & 30 stop; revocation untimed \\
\hline
eBPF containment vs.\ SIGKILL / userspace & bcc~0.29.1, bpftrace~0.20.2 & containment 30; 1000 each/run, 5 runs $\times$ 10 loads \\
\hline
Managed-cloud eviction & DOKS 3-node, k8s~API & analyzed: forced 1000; rolling 2958; preemption 1150; drain 1150; grace 552 \\
\hline
Capture-budget DoS (per-identity bucket) & CIC-DDoS + CERT/LANL replay & deterministic \\
\hline
Evidence-log scalability & \texttt{multiprocessing}, 1--12 writers & 30 runs/config (2\,s ea.) \\
\hline
Evidence-log tamper detection & SHA-256 chain + RSA-2048 anchor & 1000 trials \\
\hline
Real-container capture race & Docker + \texttt{runc}/CRIU 4.2 & 1000 cap; 200 kill \\
\hline
Sequencing barrier (container; process) & \texttt{runc}/CRIU + direct SIGKILL & 1000 each \\
\hline
Release-policy hold cost & CRIU + 64\,KiB local-write loop & 1000 each \\
\hline
Storage/coverage sensitivity analysis & Eq.~\mainref{eq:storage} model & analytic grid \\
\hline
Real-detector closed loop & Falco~0.44.1, Suricata~7.0.3 & 1000/detector \\
\hline
Memory + writable-area composition & CRIU anchor + filesystem snapshots & FS 1000/delta; CRIU anchor 1000 \\
\hline
Integrated k3s ordering demonstration & FastAPI + Cedar + FMP & 30 race; 1 chain \\
\hline
\end{tabular}
\end{table*}

\section{Robustness Analyses}
\label{sec:s-robustness}
Table~\ref{tab:s-threat-map} maps the main-paper threat objectives to the corresponding FMP mechanisms and supporting analyses. This section then expands robustness analyses summarized in Section~\mainref{subsec:feas-robustness}.

\paragraph{Weight sensitivity} The alternatives to the reference weight vector cited in Section~\mainref{subsec:feas-provenance} are, in the element order (identity, identity-attributes, action, resource, risk context, verdict, temporal-order resolution): equal ($1/7$ each), identity-heavy $(0.25, 0.25, 0.075, 0.075, 0.10, 0.10, 0.15)$, and risk-heavy $(0.10, 0.10, 0.075, 0.075, 0.40, 0.10, 0.15)$. All five measured sources and the synthetic reference were re-scored under each vector; the source ranking is identical throughout (LANL, CERT, then NSL-KDD, UNSW, CIC-DDoS). Those three vectors all retain a $0.15$ temporal-order weight, which no measured source earns, so they cannot test whether including that element depresses every empirical value. A fourth vector therefore \emph{removes} temporal-order and renormalizes the remaining six weights. Because the element scores zero for all five sources, every coverage value rises by the same factor $1/0.85{=}1.176$: CERT $64.0\to75.3$, LANL $75.0\to88.2$, NSL-KDD $30.0\to35.3$, UNSW $24.6\to28.9$, and CIC-DDoS $18.0\to21.2$. The ranking and identity-versus-network separation are unchanged (identity $75$--$88\%$ versus network $21$--$35\%$). Because all five sources score zero on temporal order, retaining its $0.15$ weight scales every reported value by $0.85$ but does not affect comparisons.


\paragraph{Remediation timing}
Table~\ref{tab:s-remediation} splits remediation at the 75\,ms bound: only direct SIGKILL ($\approx$9\,ms) and cooperating SIGTERM ($\approx$10\,ms) terminate inside the capture window ($L{=}100\%$); container \texttt{stop}, both eviction paths, and session revocation leave memory intact ($L{=}0\%$).

\begin{table}[t]
\centering
\captionsetup{font=footnotesize,skip=3pt}
\caption{Measured remediation time and memory-artifact loss $L$ at a 75\,ms capture bound.}
\label{tab:s-remediation}
\scriptsize
\setlength{\tabcolsep}{1.5pt}
\resizebox{\columnwidth}{!}{%
\begin{tabular}{|>{\raggedright\arraybackslash}p{0.52\linewidth}|r|c|}
\hline
\textbf{Remediation mechanism} & \textbf{Time to death} & \textbf{$L$} \\
\hline
direct SIGKILL (force-kill)                        & $\approx$9\,ms    & $100\%$ \\
\hline
cooperating SIGTERM (process exits)               & $\approx$10\,ms   & $100\%$ \\
\hline
container \texttt{stop} (SIGTERM, 10\,s, SIGKILL) & $\approx$10\,s    & $0\%$ \\
\hline
k8s forced eviction (\texttt{delete --force})     & $\approx$2.1\,s   & $0\%$ \\
\hline
k8s graceful eviction (30\,s grace)               & $\approx$30\,s    & $0\%$ \\
\hline
session revocation (control-plane only)           & non-destructive   & $0\%$ \\
\hline
\end{tabular}}
\par\vspace{1pt}{\scriptsize\raggedright $L$ scopes memory-resident state; session- or flow-scoped evidence may be lost by revocation while memory persists. Per-mechanism sample counts in Table~\ref{tab:s-testbed}.\par}
\end{table}

\paragraph{Storage baseline}
Table~\ref{tab:s-baseline} compares the measured-size tiered model with a decision-record-only baseline. Both retain the same projected schema coverage, but only tiered capture adds targeted session/flow and rate-capped memory artifacts, raising steady-state storage from $\approx$7.0\,TB to $\approx$46\,TB, a $\approx$6.6$\times$ increase.

\begin{table}[t]
\centering
\captionsetup{font=footnotesize,skip=3pt}
\caption{Decision-record-only baseline versus FMP-tiered capture using measured-artifact sizing.}
\label{tab:s-baseline}
\scriptsize
\setlength{\tabcolsep}{2.5pt}
\begin{tabular}{|l|c|c|}
\hline
\textbf{Evidence retained} & \textbf{record-only} & \textbf{FMP-tiered} \\
\hline
projected schema coverage & 18--75\% & 18--75\% \\
\hline
targeted session/flow artifacts & none & available \\
\hline
full-memory artifacts & none & rate-capped \\
\hline
steady-state storage & $\approx$7.0\,TB & $\approx$46\,TB \\
\hline
\end{tabular}
\end{table}

\begin{figure}[t]
\centering
\includegraphics[width=\columnwidth]{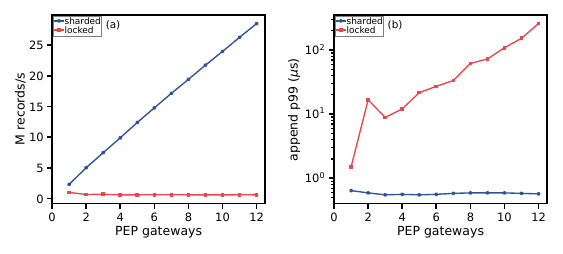}
\captionsetup{font=footnotesize,skip=3pt}
\caption{Evidence-chain append microbenchmark with $N$ writers: locked versus per-shard chains; (a) throughput, (b) append $p99$. Proposed Merkle folding is excluded from the timed path.}
\label{fig:s-eval-ledger}
\end{figure}

\paragraph{Evidence-log append throughput}
Fig.~\ref{fig:s-eval-ledger} gives the scaling behind the reported throughput values: a globally locked chain serializes at $\approx$1.0\,M records/s ($0.6$\,M contended), while per-shard chains scale near-linearly to $\approx$28\,M records/s at a flat $\approx$0.6\,$\mu$s $p99$.

\paragraph{Adaptive self-destruction}
Table~\ref{tab:s-adversarial} shows that the \texttt{memset} scrub ($\approx$3\,ms) and self-exit ($\approx$7\,ms) finish before the 65\,ms reactive capture ($L{=}100\%$); the Python overwrite ($\approx$67\,ms) overlaps the capture distribution.

\begin{table}[t]
\centering
\captionsetup{font=footnotesize,skip=3pt}
\caption{Adaptive self-destruction relative to a 65\,ms reactive capture ($n{=}1000$/strategy).}
\label{tab:s-adversarial}
\scriptsize
\setlength{\tabcolsep}{2.5pt}
\begin{tabular}{|l|r|r|c|}
\hline
\textbf{Self-destruction} & \textbf{median} & \textbf{$p95$} & \textbf{Outcome} \\
\hline
self-exit & $\approx$7\,ms & $\approx$15\,ms & $L{=}100\%$ \\
\hline
\texttt{memset} scrub & $\approx$3\,ms & $\approx$4\,ms & $L{=}100\%$ \\
\hline
Python overwrite & $\approx$67\,ms & $\approx$69\,ms & overlap \\
\hline
\end{tabular}
\end{table}

\section{Experimental Environment and Protocols}
\label{sec:s-testbed}
Tables~\ref{tab:s-nodes} and~\ref{tab:s-testbed} report the hosts and per-experiment tooling, replication, and scope. Both bare-metal hosts ran OpenSSL~3.0.13 (FIPS off), SHA-256 via \texttt{sha\_ni}, and software RSA-2048 (no TPM/HSM). Iteration counts were $2{\times}10^4$/$5{\times}10^4$/$300$/$5{\times}10^3$ for canonicalization/hash linking/RSA/flow serialization, respectively. Each evidence-log throughput point is the median of 30 two-second runs; percentiles cover $\sim$$10^4$--$10^6$ sampled appends per run (1/64 sampling).

DOKS provides only orchestrator-eviction and detector measurements; node-level checkpointing was unavailable. Attempted/initially retained counts were 1000/1000 forced, 3731/3000 rolling replacement, 1154/1150 preemption, 1152/1150 node drain, and 554/552 grace expiry. A timing-consistency check excluded 42 rolling records below the 1\,s minimum ($3.1$--$176.1$\,ms), leaving $n{=}2958$; raw and filtered medians were $2097.2$ and $2097.5$\,ms, and original records are retained. Most exclusions were pods that died before the trigger; other modes excluded 0--4 records. The criterion concerns trigger validity, not measured delay. Sessions record the test-program source hash in \texttt{config.source\_sha256\_16}. Four hashes occur, with stable medians across them (rollout $2096$--$2099$\,ms, drain $2194$--$2196$\,ms, preemption $2194$--$2199$\,ms). This supports pooled mechanism-level summaries, not ranking within $\approx$2.1--$2.2$\,s. CRIU checkpoints used \texttt{criu dump --shell-job --leave-running}; anticipatory pre-dump added \texttt{--track-mem} with \texttt{--prev-images-dir}.

\paragraph{Replay protocol} CERT, UNSW-NB15 and CIC-DDoS2019 are
seed-shuffled ($\mathrm{seed}{=}7$); NSL-KDD is replayed in native file order; and
LANL reservoir-samples benign events to the uniform $125{,}000$-event target
while retaining every red-team event. Timestamps are consumed where the adapter reads them: LANL
uses the native one-second epoch field, CERT parses the record date to an epoch
value, and NSL-KDD, UNSW-NB15 and CIC-DDoS2019 receive a row-index sequence
instead. For NSL-KDD and UNSW-NB15 the evaluated datasets expose no wall-clock
field; for CIC-DDoS2019 the flow records do carry a timestamp that this adapter
does not consume, which is an adapter limitation rather than a dataset property.
The row-index value is a deterministic ordering surrogate only: it is not
seconds, and no rate, interval, or burst statistic is derived from it.

Replay order and this surrogate therefore do not enter the capture-budget
calculation. The malicious-event rate is $\lambda_m{=}hR$, where $h$ is the post-sampling labeled fraction, not an estimate of native prevalence; $R{=}500$/s is the stipulated operating point of Section~\mainref{subsec:feas-cost}, not a rate recovered from the data.
The quantity $\min(1,c/\lambda_m)$ is thus an admission ceiling under a
stationary offered load at that operating point: it bounds the fraction of
eligible malicious events a $c{=}0.05$/s budget can admit, and assumes perfect
admission among them with no false positives. Burstiness is not modeled in this replay;
Section~\ref{sec:s-cost} gives a separate sensitivity calculation. Schema coverage is a field-presence
projection computed per event, so neither replay order nor the timestamp
surrogate affects it, which is why the CIC-DDoS adapter limitation leaves the
reported coverage values unchanged.

\paragraph{Capture-budget DoS protocol} The admission replay merges a CIC-DDoS2019
flood ($63{,}708$ high-risk events across $2$ addresses mapped to one adversarial identity in the concentrated condition) with the labeled CERT or LANL incident stream over a $3600$\,s window: $1199$ legitimate incidents
across $K{=}66$ identities for CERT, $702$ across $K{=}98$ for LANL. Admission is
two-level. A global token bucket refills at the full-memory cap $c{=}0.05$/s, and
each identity holds its own bucket refilling at the fair share $c/(K{+}1)$, which
is $0.000746$/s for CERT and $0.000505$/s for LANL; the extra identity in the
denominator is the adversary. The global bucket has capacity $4$ and starts full;
each identity bucket is created full on first appearance. A first request satisfies
the identity bucket but still requires a global token. A
capture is granted only when both levels hold a token, and a token is charged at
neither level on a denial. The baseline condition runs the same code path with
the per-identity level disabled, so baseline and treatment differ only in that
one flag.

\paragraph{Evidence-log benchmark scope} The throughput results time a single-host, in-memory hash-chain append with $N$ PEP writers as $N$ processes. They exclude network transport, durable writes, and Merkle folding. In the proposed design, a coordinator would anchor shard heads asynchronously into a global root off the append path. The reported rates characterize only the in-memory chaining primitive, not a deployed evidence store. Tamper detection is a separate 1000-trial experiment against a $10{,}000$-entry evidence log anchored every $1000$ records. Each trial alters a stored field, deletes an entry, or replaces a stored principal value, then tests whether the next anchor exposes the change. Detection is claimed only for
alteration after anchoring and only while the signing key and evidence root are
uncompromised.

\begin{table}[t]
\caption{Matched A--D study provenance. Primary study 2026-08-22.}
\label{tab:s-ad-provenance}
\centering\scriptsize
\setlength{\tabcolsep}{3pt}
\begin{tabularx}{\columnwidth}{|>{\raggedright\arraybackslash}p{1.9cm}|X|}
\hline
\textbf{Host} & Zen~5 \texttt{avi-TRX50-AERO-D}, kernel \texttt{7.0.0-29-generic}, governor \texttt{performance} \\
\hline
\textbf{Tooling} & CRIU~4.2, runc~1.3.4, Docker~29.1.3 \\
\hline
\textbf{CRIU config} & \texttt{default.conf} frozen at \texttt{5e0bf1d4\ldots}, recorded per trial; a differing hash invalidates it. A--C run with \texttt{runc.conf} absent; D installs a pinned \texttt{track-mem} file under lock, removed with fail-closed verification. Captures pass \texttt{--tcp-established --ext-unix-sk --file-locks} explicitly \\
\hline
\textbf{Alert correlation} & Unique container per trial; an event is accepted only with that exact id \emph{and} name and \texttt{evt.time} strictly after the trigger and inside the trial window. Events without identity are rejected; accepted and rejected near-matches are retained. Minimum lead $+0.42$\,s over 1000 trials, with zero pre-trigger, cross-container, missing-identity or ambiguous acceptances \\
\hline
\textbf{Cadence} & Absolute monotonic deadlines; observed $2.000$/$1.000$\,s ($0.00\%$ error). Per-snapshot start, frozen interval, completion, parent-chain identity recorded \\
\hline
\textbf{Statistical unit} & Clopper--Pearson for rates, Newcombe--Wilson for differences. D reduces each trial to one predeclared summary before bootstrapping, so the trial is the unit of replication; a cluster bootstrap is a sensitivity check \\
\hline
\textbf{Counts} & $200$ each, $1000$ total, all valid; sample size fixed in advance; no replacement \\
\hline
\textbf{Not pooled} & Two retained preliminary studies: one lacking freeze and containment-completion metrics, one whose correlation accepted pre-trigger events \\
\hline
\end{tabularx}
\end{table}

\paragraph{Measurement protocol} Primary timings use the Zen~5 host (\texttt{ext4}/NVMe, \texttt{performance} governor, no core isolation); EPYC/A100 supports replay, behavioral end-to-end measurements, and capture cross-checks (Table~\ref{tab:s-nodes}). Latencies use \texttt{perf\_counter} from trigger to completion; the enforcement study is pinned to four cores. Gateway overhead uses one uvicorn worker, concurrency~64, 3000 warm-up then $3{\times}10^4$ timed requests each, and 30 paired repeats (Table~\mainref{tab:overhead}). The evidence log links $h_i{=}\mathrm{SHA\text{-}256}(\mathit{rec}_i\,\|\,h_{i-1})$ and signs an RSA-2048 anchor every 1000 entries ($n{=}10^4$, 1000 trials); percentiles use $1/64$ sampling over each two-second run (30-run median). Replays are deterministic under fixed seeds. Falco alerts or a custom Suricata rule for off-host C2 trigger a \texttt{runc} CRIU checkpoint (\texttt{--tcp-established --leave-running}). Recovery endpoints are the four planted memory markers and the \textsc{established} C2 socket (Table~\mainref{tab:detectors}). Only the duration summary uses each trial's \emph{terminating} checkpoint; intermediate misses remain failures for the preregistered first-attempt endpoint.

\renewcommand{\IEEEbibitemsep}{0pt plus 0.1pt}

\renewcommand{\IEEEbibitemsep}{0pt plus 0.1pt}
\balance
\bibliographystyle{IEEEtran}
\bibliography{references}

\end{document}